\documentclass[11pt,a4paper]{article}

\usepackage{amssymb,amsmath} 
\usepackage{soul}
\usepackage{amsthm} 
\usepackage{color}
\usepackage{subcaption}

\usepackage{lineno}    

\usepackage{authblk}
\usepackage{color,graphicx}

\newcommand{\I}{\mathrm{i}}
\newcommand{\D}{\mathrm{d}}
\newcommand\myeq{\stackrel{\text{\tiny def}}{=}}

\title{Estimation of spatio-temporal wave grouping properties using Delaunay triangulation and spline techniques}
\date{}

\author[1]{Jos\'e Carlos Nieto Borge\thanks{Email: josecarlos.nieto@uah.es}}
\author[1]{Juan Gerardo Alc\'azar\thanks{Email: juange.alcazar@uah.es}}
\author[1]{David Orden\thanks{Email: david.orden@uah.es}}
\author[1]{Sara Marazuela Reca}
\author[1]{Gerardo Rodr\'{\i}guez}

\affil[1]{Department of Physics and Mathematics. Universidad de Alcal\'a, 28805 Alcal\'a de
Henares, Madrid, Spain.}

\begin{document}
\maketitle

\begin{abstract}
Wave groups can be detected and studied by using the \emph{wave envelope}. So far, the method used to compute the wave envelope employs the Riesz transform. However, such a technique always produces symmetric envelopes, which is only realistic in the case of linear waves. In this paper we present a new method to compute the wave envelope providing more realistic results. In particular, the method allows to detect non-symmetry in the wave envelope, something useful, for instance, when detecting groups of high waves. The method computes first the local maxima and minima of the sea surface, and then determines the wave envelope by combining discrete methods, namely the use of the Delaunay triangulation, and tensor-product splines. The proposed method has been applied to simulated wave fields, and also to wave elevations data measured by an X-band radar. The obtained results correctly reproduce the behavior of the simulated waves.
\end{abstract}


\section{Introduction}
\label{sc:Introduction}

Ocean gravity waves propagate in the ocean as packages of consecutive high waves traveling together~\cite{Longuet-Higgins1984,Longuet-Higgins1986}. This phenomenon is known in the literature as {\it wave grouping}~\cite{Ochi2005,Goda2010}.
Those wave groups are specially dangerous for marine activities, such as navigation, coastal management, on- and off-shore platform operation, etc.
This danger is not only caused by the presence of high waves, but also by the fact that those waves propagate with nearly equal periods, which can produce stability problems to marine structures when the wave periods involved in the groups are close to the resonant period of the structure, such as moving vessels, breakwaters, etc.~\cite{Ochi2005,Claussetal2008}.

Although the analysis of wave grouping has been carried out for several decades \cite{Hamiltonetal1979,MolloChristensen+Ramamonjiariosa1980,Kimura1980}, this phenomenon and its properties are not fully explained yet \cite{Ochi2005}, and, therefore, it needs to be analyzed in detail for a better understanding of the dynamical mechanisms involved in the wave group propagation, such as the persistence of the groups, number of waves within a group, etc.

Wave grouping has been studied traditionally in the temporal domain $t$ by using wave elevation time series acquired by in situ-sensors (e.g.,~anchored buoys, pressure gauges, wave lasers, etc.).
Hence, those measurements can characterize  wave grouping features at a fixed position (i.e., the point of deployment) \cite{Medina+Hudspeth1990,Donelanetal1996}.
Alternatively to those point measurements, in recent years different remote imaging techniques capable of studying the sea surface in space and time have been developed.
Some of these techniques are based on the use of passive sensors, like video cameras \cite{Piotrowski+Dugan2002,Gallegoetal2011,Benetazzoetal2012,Ruttenetal2017}.
Other techniques use active microwave sensors, such as incoherent and coherent radars mounted on off- and on-shore stations or moving vessels \cite{Ziemer+Dittmer1994,Buckleyetal1994,Buckley+Aler1997,Buckley+Aler1998,Nieto-Borge04,Trizna2001,Ziemeretal2004,Dankert+Rosenthal2004b,Belletal2005,Wuetal2011,Stole-Hentschel18}.
All those remote sensing techniques are able to acquire temporal sequences of images of
the sea surface.
The spatio-temporal information provides a more detailed description of the wave elevation properties than the historical measurements based on time series analysis and, therefore, it requires the use of additional tools, or even new techniques of analysis, to obtain information of the wave field evolution in space and time \cite{Lietal2014}.

In one-dimensional records (i.e., wave elevation data depending on only one parameter, such as heave time series measured by a buoy), the individual wave heights can be estimated by applying the zero-up crossing method, or similar techniques, but those methods are difficult to extend to a higher number of dimensions.
A way to study the spatio-temporal evolution of wave heights and groups consists of the estimation of the wave envelope depending on the variables $(x, y, t)$.
For linear and narrow-banded wave fields the envelope can be estimated by the Riesz Transform (RT) \cite{Nieto-Borge13}, which is a multidimensional generalization of the Hilbert transform (HT)\cite{King2009}.
For nonlinear and$/$or broadband wave fields RT, or HT for time series, is not the appropriate solution.
The reason is that the RT always leads symmetry between the upper and the lower envelopes \cite{King2009,Nieto-Borge13}.
Furthermore, the narrowband hypothesis reduces the irregularity of the envelope passing through most of the local maxima, and the local minima for the case of the lower envelope \cite{Ochi2005}.
 This is accurate for linear wave fields, but not for non-linear wave fields, where there may be asymmetries between
wave crests and wave troughs with respect to the mean sea level.
We argue this in more detail in Section 2.

Therefore, other techniques should be applied. An example of an alternative technique appears in \cite{Sanina15}, where the authors reconstruct the two-dimensional envelope of nonlinear wave fields by applying sets of one-dimensional splines parallel to the mean wave propagation direction in the $(x, y)$-domain for different time steps $t$. The approach that we present in this paper also makes use of splines, but proceeds in a completely different way to that in \cite{Sanina15}. The main idea is to use two-dimensional spline interpolation, to refine a first (linear) approximation of the envelope computed by using discrete methods, namely Delaunay triangulation \cite{deBerg2008,Fortune2004}.

In more detail,
for a fixed time $t$ our method first locates the wave points corresponding to local positive maxima and local negative minima of the wave elevation function, by comparing the elevation of each point with the elevations of its immediate neighbours. Then, we use the Delaunay triangulation to build a piecewise-linear model of the envelope. Finally, we refine this model by using tensor-product spline interpolation
\cite{Lyche-Morken08,Patrik02}
This refinement step is necessary, because the linear approximation introduces fictitious harmonics and spectral noise, which however disappear when the spline refinement is applied. Therefore, while the method in \cite{Sanina15} uses one-dimensional splines to build slices of the wave envelope surface, we directly build the surface in one go, considering the two-dimensional structure of the sea surface.

The method proposed in this paper is suitable both for linear and non-linear wave fields, and has been used for different simulated wave elevation fields $\eta(x, y, t)$ using the standard stochastic approach of linear wave fields and a second-order nonlinear approach based on the model proposed by \cite{Tayfun80} \cite{Nieto-Borge05}.

The paper is structured as follows:
Section~\ref{sc:Envelope} describes the properties of the envelope derived from RT for linear wave fields.
Section~\ref{sc:MathematicalMethods} describes the mathematical techniques used in the method, namely triangulations and tensor-product spline interpolation, as well as the method itself. Section~\ref{sc:Experiments} deals with a brief description of the wave field simulation methods (i.e. linear and second-order approach stochastic models) used in this paper, as well as the corresponding results obtained for the estimation of the related wave grouping properties. Section~\ref{sc:additional_observations} addresses the effect of measurement errors in the method.
In Section \ref{sc:new}, the method is applied to a wave elevation map measured by an X-band radar mounted at the German platform FINO 1, located on the North Sea.
Finally, Section~\ref{sc:Conclusions} summarizes our conclusions.

\section{Spatio-temporal wave envelope for linear and narrow-banded wave fields}
\label{sc:Envelope}

Under the frame of the linear wave theory, the wave elevation field $\eta({\bf r}, t)$, where ${\bf r} = (x, y)$, is regarded as a superposition of different monochromatic wave components \cite{Ochi2005}, where each component is characterized by its amplitude $a$, wave number vector ${\bf k} = (k_x, k_y)$, angular frequency $\omega$, and phase $\varphi$. Hence, using a discrete notation, $\eta({\bf r}, t)$ can be expressed as

\begin{equation}\label{eq:SeaState}
\eta({\bf r}, t) = \sum_m a_m \cos \left( {\bf k}_m \cdot {\bf r} - \omega_m t + \varphi_m \right)
\mbox{.}
\end{equation}

\noindent
Eq. \ref{eq:SeaState} corresponds to an Eulerian description of the wave elevation process~$\eta$ \cite{Krogstad+Trulsen2010},
where wave numbers~${\bf k}$ and frequencies~$\omega$ are dependent through the dispersion relation for linear gravity waves (i.e., $\omega = \sqrt{g k \tanh (k h)}$, being $k = |{\bf k}|$, $g$ the acceleration of the gravity and $h$ the water depth).
Those wave fields given by \eqref{eq:SeaState} are considered as zero-mean Gaussian stochastic processes, where the spectral components are statistically independent, being $a_m$ and $\varphi_m$ random variables, which are statistically homogeneous in space and stationary in time \cite{Goda2010}.
Under these conditions, and taking into account \eqref{eq:SeaState}, the variance of $\eta$ is

\begin{equation}\label{eq:VarianceEta}
\sigma^2 =
\mathcal{E}\left[ \eta^2 \right] =
\frac{1}{2} \sum_m \mathcal{E}\left[ a_m^2 \right]
\mbox{,}
\end{equation}

\noindent
where $\mathcal{E}\left[ \cdot \right]$ denotes the expectation operator. Note that Eq. \ref{eq:VarianceEta} considers $\mathcal{E}\left[ \eta \right] = 0$.
As Eq. \ref{eq:SeaState} represents a Gaussian process, this model describes wave fields with statistical symmetry between wave crests and troughs \cite{Ochi2005}.


\subsection{Spatio-temporal estimation of the wave envelope by using the Riesz Transform}
\label{sc:RieszTrnsfm}

In a similar way that is done for wave elevation time series \cite{Goda2010}, assuming that the process $\eta$ is linear narrow-banded \cite{Ochi2005},
the spatio-temporal description of linear wave elevation fields $\eta({\bf r}, t)$ can be factorized using the so-called local and instantaneous amplitude $A({\bf r}, t)$ and local and instantaneous phase $\Phi({\bf r}, t)$ as \cite{Nieto-Borge13}

\begin{equation}\label{eq:APhiFactorization}
	\eta({\bf r}, t) = A({\bf r}, t) \cos \Phi({\bf r}, t)
	\mbox{.}
\end{equation}

\noindent
The spatio-temporal wave grouping properties are described from the local and instantaneous amplitude $A({\bf r}, t)$ \cite{Nieto-Borge13}.
In the development of \eqref{eq:APhiFactorization}, it is assumed that the wave field is described by a characteristic wave number and frequency, which can be understood as the carrier of the process $\eta$ \cite{Goda2010}.
This fact is a consequence of the narrow-band approach and the assumption of the linear wave theory \cite{Ochi2005}.
Furthermore, \eqref{eq:APhiFactorization} indicates that $- A \le \eta \le A$, being $- A({\bf r}, t)$ and $A({\bf r}, t)$ the so-called lower and upper envelope respectively \cite{Nieto-Borge13}.
Those magnitudes are denoted in this paper as $A^{-} \equiv - A$, and $A^{+} \equiv A$.
It can be seen that the model described by \eqref{eq:APhiFactorization} has symmetric lower and upper envelopes (i.e. $A^{-} = - A^{+} = - A$), This is a consequence of the linearity assumption but not of the narrowband approach. Under these conditions, the wave height is regarded as twice the amplitude $H \sim 2 A$ \cite{Goda2010}.
However, for non-linear wave fields, where there may be asymmetries between wave crests and wave troughs with respect to the mean sea level, the upper and lower envelopes are not symmetric either (i.e. $A^{-}({\bf r}, t) \neq - A^{+}({\bf r}, t)$). For those cases, equation \eqref{eq:APhiFactorization} has some limitations to describe the wave grouping phenomenon.

For those cases where the hypothesis of \eqref{eq:APhiFactorization} holds, the local and instantaneous wave envelope $A({\bf r}, t)$ may be estimated by the Riesz Transform (RT) \cite{Nieto-Borge13}, which is a multidimensional generalization of the Hilbert Transform used for wave grouping analysis of wave elevation time series \cite{Medina+Hudspeth1990}.
For a given time $t$, the two-dimensional RT of the wave elevation field $\eta({\bf r}, t)$ is defined as \cite{King2009}

\begin{equation}\label{eq:DefRieszT}
\hat{\eta}_{j}({\bf r}, t) =\frac{1}{2 \pi}  \lim_{\varepsilon \to 0} \int_{\left| \boldsymbol{\lambda} \right| > \varepsilon} \eta({\bf r}, t) \frac{x_j - \lambda_j}{\left| {\bf r} - \boldsymbol{\lambda} \right|^{3}}  \D \lambda_1 \D \lambda_2
\quad \mbox{,} \qquad
j = 1 , \, 2
\mbox{ ,}
\end{equation}

\noindent
where $\boldsymbol{\lambda} =(\lambda_1, \lambda_2)$.
The index $j$ takes the values $j=1$ or $j=2$ denoting the directions $x \equiv x_1$ and $y \equiv x_2$ respectively.
Note that RT has two components in this particular case, $\hat{\eta}_{1}({\bf r}, t) \equiv \hat{\eta}_{x}({\bf r}, t)$, and $\hat{\eta}_{2}({\bf r}, t) \equiv \hat{\eta}_{y}({\bf r}, t)$.
From the two components of RT a vector field $\boldsymbol{\eta}({\bf r}, t)$ is constructed as \cite{King2009}

\begin{equation}\label{eq:VectorRT}
\boldsymbol{\eta}({\bf r}, t) = \left( \eta({\bf r}, t),  \hat{\eta}_{x}({\bf r}, t), \hat{\eta}_{y}({\bf r}, t) \right)^{T}
\mbox{.}
\end{equation}

\noindent
The estimation of the local and instantaneous amplitude $A({\bf r}, t)$ is obtained as the norm of the vector field $\boldsymbol{\eta}({\bf r}, t)$

\begin{equation}\label{eq:AmplitudeVectorRT}
A({\bf r}, t) = \left| \boldsymbol{\eta}({\bf r}, t) \right| = \sqrt{\eta^2({\bf r}, t) + \hat{\eta}_{x}^2({\bf r}, t) + \hat{\eta}_{y}^2({\bf r}, t)}
\mbox{.}
\end{equation}

\noindent
The respective RT estimation of the local and instantaneous phase $\Phi_r({\bf r}, t)$ is given by

\begin{equation}\label{eq:PhiRT}
\Phi({\bf r}, t) = \tan^{-1} \left[ \frac{ \sqrt{\hat{\eta}_{x}^2({\bf r}, t) + \hat{\eta}_{y}^2({\bf r}, t)}}{\eta({\bf r}, t)} \right]
\mbox{.}
\end{equation}

\noindent
From \eqref{eq:AmplitudeVectorRT} and \eqref{eq:PhiRT}, the wave elevation field $\eta$ is derived as an expression equivalent to \eqref{eq:APhiFactorization}.

\noindent
For practical cases, \eqref{eq:DefRieszT} is not applied and RT is computed by using the relationship between the Fourier transforms of the original signal $\eta$ and its respective RT vector components \cite{King2009,Nieto-Borge13}

\begin{equation}\label{eq:FourierRT}
\mathcal{F}\left[\hat{\eta}_{j} \right] = - \I \frac{k_j}{\left| {\bf k} \right|} \mathcal{F}\left[\eta \right]
\quad \mbox{;} \qquad
j = 1 , \, 2
\mbox{;}
\end{equation}

\noindent
where $\mathcal{F}$ denotes the two-dimensional Fourier Transform from the spatial to the wave number domain (${\bf r} \rightarrow {\bf k}$) for a given fixed time $t$, while
$k_1$ and $k_2$ are the two components of the wave number vector ${\bf k} =(k_x, k_y)\equiv (k_1, k_2)$.
Taking into account the linear wave field $\eta({\bf r}, t)$ described by \eqref{eq:SeaState}, the two components of RT given by \eqref{eq:FourierRT} are

\begin{eqnarray}
\hat{\eta}_{x} ({\bf r}, t)
& = & \sum_m a_m \cos \theta_m \sin \left( {\bf k}_m \cdot {\bf r} - \omega_m t + \varphi_m \right)
\mbox{,} \label{eq:SeaStateRTx} \\
\hat{\eta}_{y} ({\bf r}, t)
& = & \sum_m a_m \sin \theta_m \sin \left( {\bf k}_m \cdot {\bf r} - \omega_m t + \varphi_m \right)
\mbox{.} \label{eq:SeaStateRTy}
\end{eqnarray}

\noindent
where $\theta_m = \arg {\bf k}_m$ is the propagation direction of the $m^{\rm th}$ spectral component.
Therefore,
$\cos \theta_m = k_{x_m} / \left| {\bf k}_m \right|$, and
$\sin \theta_m = k_{y_m} / \left| {\bf k}_m \right|$,
which are related with the transfer functions of RT given by Equation \ref{eq:FourierRT}. Equations
\ref{eq:SeaStateRTx} and~\ref{eq:SeaStateRTy} are the horizontal Lagrangian wave displacements at the mean sea level ($z = 0$) of the wave elevation field given by Eq. \ref{eq:SeaState} \cite{Krogstad+Trulsen2010,Nieto-Borge13}.
Hence, the local and instantaneous wave envelope can be understood as the norm of the vector field $\boldsymbol{\eta}({\bf r}, t)$, whose components are the wave elevation and the horizontal wave displacements in a single wave cycle.
The expectation of $A^2$ defined in \eqref{eq:AmplitudeVectorRT} is twice the variance of $\eta$

\begin{equation}\label{eq:MeanA2}
	\mathcal{E}\left[ A^2 \right] =
	\mathcal{E}\left[ \eta^2 \right] +
	\mathcal{E}\left[ \hat{\eta}^2_{x} \right] +
	\mathcal{E}\left[ \hat{\eta}^2_{y} \right] = 2 \sigma^2
	\mbox{,}
\end{equation}

\noindent
where \eqref{eq:VarianceEta} has been used, as well as
\eqref{eq:SeaStateRTx} and \eqref{eq:SeaStateRTy}
for the variances:
$\mathcal{E}\left[ \hat{\eta}^2_{x} \right] = \sigma^2 / 2$, and
$\mathcal{E}\left[ \hat{\eta}^2_{y} \right] = \sigma^2 / 2$.

\subsubsection{Spectral components of the envelope derived from RT}
\label{sc:SpectrCompEnvelope}

The location of the spectral components of the envelope $({\bf k}_A, \omega_A)$ can be identified considering $A^2$ in \eqref{eq:AmplitudeVectorRT} together with \eqref{eq:SeaState},
\eqref{eq:SeaStateRTx}, and \eqref{eq:SeaStateRTy} for $\eta^2$, $\hat{\eta}^2_{x}$, and $\hat{\eta}^2_{y}$ respectively. Thus,

\begin{equation}\label{eq:A2G2P2}
	A^2({\bf r}, t) = G^2({\bf r}, t) + P^2({\bf r}, t)
	\mbox{,}
\end{equation}

\noindent
where, using the definition $\Phi_m \equiv {\bf k}_m \cdot {\bf r} - \omega_m t + \varphi_m$, the terms
$G^2({\bf r}, t)$ and $P^2({\bf r}, t)$ are given by

\begin{equation}\label{eq:G2Def}
	G^2({\bf r}, t) =
	\frac{1}{2} \sum_m\sum_n a_m a_n \left[1 + \cos (\theta_m - \theta_n) \right] \cos \left( \Phi_m - \Phi_n \right)
	\mbox{,}
\end{equation}

\begin{equation}\label{eq:P2Def}
	P^2({\bf r}, t) =
	\frac{1}{2} \sum_m\sum_n a_m a_n \left[1 - \cos (\theta_m - \theta_n) \right] \cos \left( \Phi_m + \Phi_n \right)
	\mbox{.}
\end{equation}

\noindent
From \eqref{eq:G2Def} and \eqref{eq:P2Def}, it can be seen that the spectral components of $G^2({\bf r}, t)$ correspond to second-order differences of the spectral variables, ${\bf k}_A = {\bf k}_m - {\bf k}_n$, $\omega_A = \omega_m - \omega_n$, while $P^2({\bf r}, t)$ has second-order summation components ${\bf k}_A = {\bf k}_m + {\bf k}_n$, $\omega_A = \omega_m + \omega_n$.
Therefore, $G^2({\bf r}, t)$ is responsible for the long scale spatio-temporal evolution (long distances and slow times), and $P^2({\bf r}, t)$ of the short scale spatio-temporal changes (short distances and fast times).
The magnitude $G^2$ is commonly known in the literature as group train, and $P^2$ as pulse train \cite{Gran1992}.
Taking into account that the spectral components of the wave field $\eta$ are statistically independent, the expectation operator applied to \eqref{eq:G2Def} and \eqref{eq:P2Def} is

\begin{eqnarray}
	\mathcal{E} \left[ G^2 \right] & = & \sum_m \mathcal{E} \left[ a_m^2 \right] = 2 \sigma^2 \mbox{,} \label{eq:EnergyG2} \\
	\mathcal{E} \left[ P^2 \right] & = & 0   \mbox{.} \label{eq:EnergyP2} \\
\end{eqnarray}

\begin{figure}[ht]
\begin{center}
  \includegraphics[width=\textwidth]{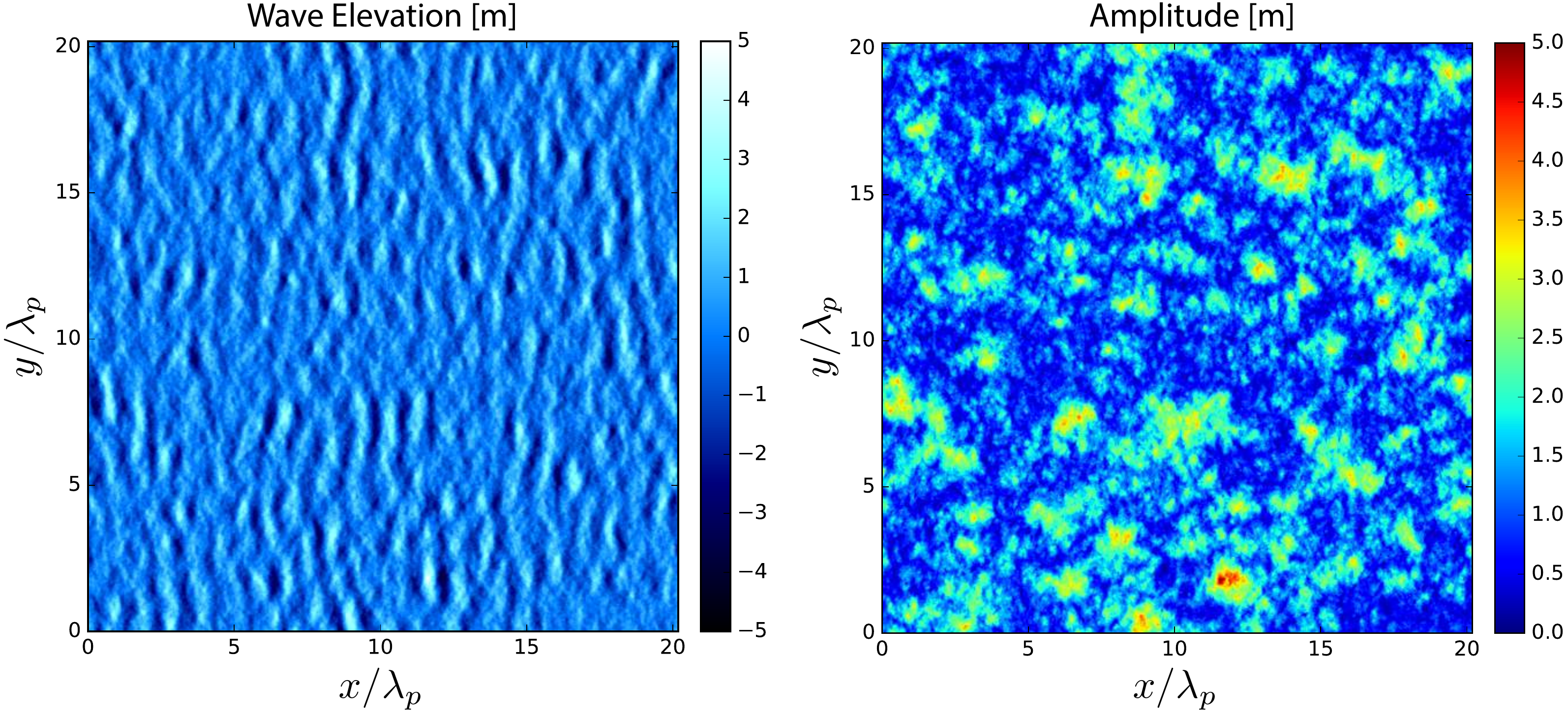}
\end{center}
\caption{Simulated linear wave field (left) and the corresponding estimation of the wave envelope derived from RT (right).
The simulated wave field corresponds to a JONSWAP spectrum case with significant wave height $H_s =4$~m, peak period $T_p = 10$~s, and peak wave length derived from the wave number spectrum $\lambda_p = 158$~m. The directional spreading parameters used in this simulation where $s_{max} = 15$, $\mu_1 = 5$, and $\mu_2=-2.5$.}\label{fg:LinSimExample}
\end{figure}

\begin{figure}[ht]
\begin{center}
\includegraphics[width=\textwidth]{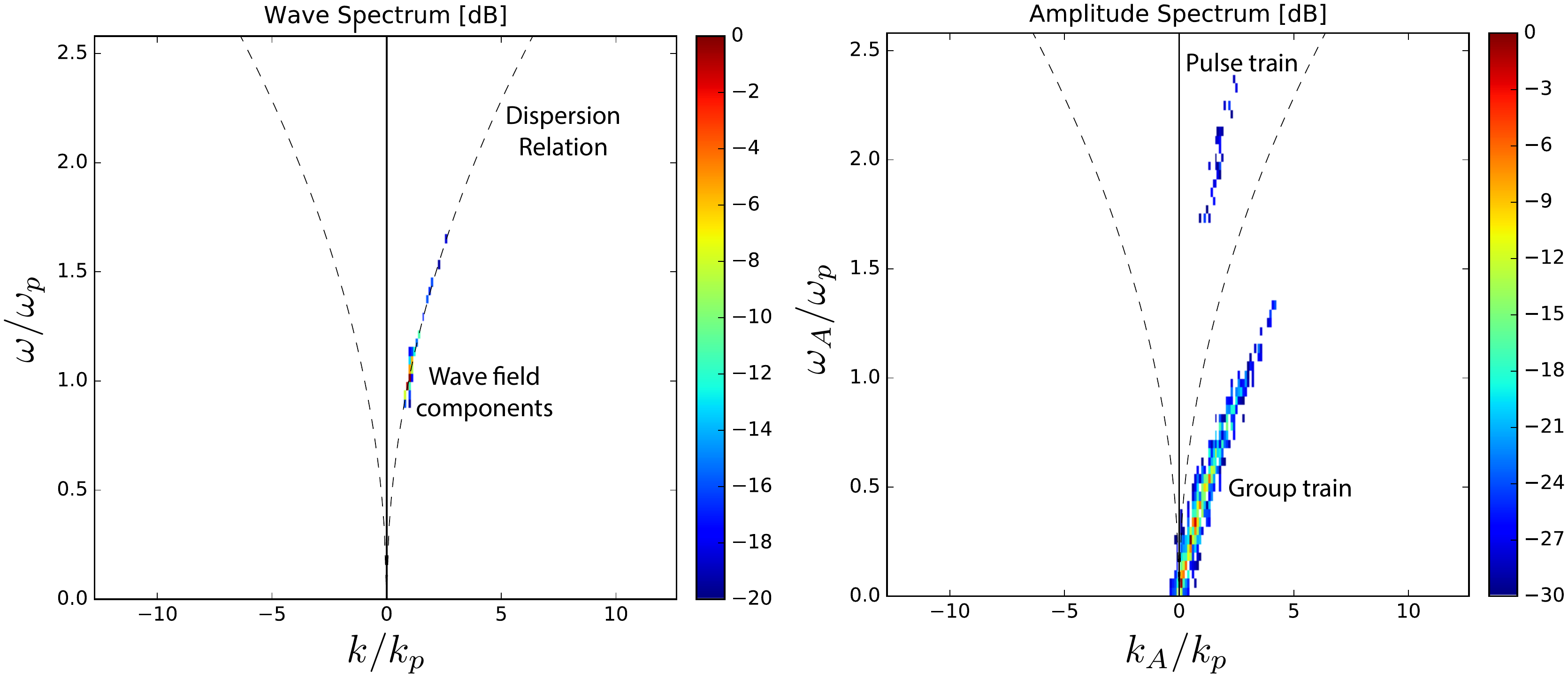}
\end{center}
\caption{Wave spectrum (left) and envelope spectrum (right) of the example shown in Figure~\ref{fg:LinSimExample}.
Both spectral representations correspond to a transect along the mean wave propagation direction in the $({\bf k}, \omega)$-domain.
The spectral components of the group train~$G^2$ and the pulse train~$P^2$ can be identified in the region of low and high wave numbers and frequencies respectively.
The spectral variables are normalized to the peak wave number~$k_p$ and peak frequency~$\omega_p$ of the wave field.
The dashed line corresponds to the dispersion relation $\omega({\bf k})$.}\label{fg:LinSimExampleSpectra}
\end{figure}

\noindent
The group train $G^2$ is responsible of the wave energy propagation as the mean value of the pulse train $P^2$ vanishes \cite{Nieto-Borge13}.
Assuming the existence of a dispersion relation $\omega({\bf k})$, the ``phase speed'' of the spectral component $({\bf k}_A, \omega_A)$ of $G^2$ is

\begin{equation}\label{eq:GroupVelocityG2}
   \frac{\omega_A}{|{\bf k}_A|}	=
    \frac{\omega_m - \omega_n}{|{\bf k}_m - {\bf k}_n|} =
   \frac{\omega({\bf k}_m) - \omega({\bf k}_n)}{|{\bf k}_m - {\bf k}_n|}
   \xrightarrow[{\bf k}_n \to {\bf k}_m]{}
   \frac{\D \omega}{\D k}
   \mbox{,}
\end{equation}

\noindent
which is the group velocity of the wave field components.
As an example, Fig.~\ref{fg:LinSimExample} shows a simulated linear wave field $\eta$ (left) depending on the sea surface coordinates ${\bf r}$ at a fixed time $t$, and its amplitude estimation $A$ derived from RT (right). Their wave number and frequency spectra are illustrated in Fig.~\ref{fg:LinSimExampleSpectra}. The spectrum of the envelope can be seen on the right part of that figure, where the spectral components of the group and pulse trains can be identified.

Taking into account \eqref{eq:EnergyG2} and \eqref{eq:GroupVelocityG2}, the relevant magnitude to analyze the spatio-temporal properties of wave grouping is the group train~$G^2$ rather than the pulse train~$P^2$.
Therefore, as $P^2({\bf r}, t)$ evolves in very short scales of distances and times, its contribution adds noise in the envelope for the wave grouping study.
This effect is well known for time series analysis of wave groups derived from buoy records \cite{Longuet-Higgins1984,Longuet-Higgins1986}.
In that case, the estimation of the envelope derived from HT is applied to wave elevation time series previously filtered with a band-pass filter centered in the peak frequency $\omega_p$.
This method is difficult to generalize to higher number of dimensions, as the $({\bf k}, \omega)$-domain.
Therefore, an alternative estimation method of the spatio-temporal envelope, without the need of applying any kind of additional filter, would be more useful. In addition, as it has already been mentioned, RT provides symmetric upper and lower envelopes, which are not representative of linear or not-narrow-banded wave fields.
Thus, for any kind of wave field, it is necessary to use different approaches that permit the estimation of independent smooth upper $A^{+}({\bf r}, t)$ and lower envelope $A^{-}({\bf r}, t)$ that pass through all the local maxima (crests) and minima (troughs) respectively.
The following section deals with the description of the techniques proposed in this paper to estimate the envelopes, based on discrete and spline methods.

\section{Estimation of wave envelopes using discrete and spline methods}
\label{sc:MathematicalMethods}

In this section we present an alternative mathematical method to construct the upper and lower wave envelopes. Here we assume that we have measured the wave elevations $\eta({\bf r}, t_0)$ at a given time $t=t_0$, at several points ${\bf r}_{mn}=(x_m,y_n)=(m\Delta x,n\Delta y)$, with $m=0,1,\ldots,N_x-1$ and $n=0,1,\ldots N_y-1$, of a rectangular grid, see Fig.~\ref{ElevationNeighbors}. Furthermore, we will denote by $\eta_{mn}$ the elevation at the point ${\bf r}_{mn}$. The method consists of three steps (we will later refer to these steps as Step 1, Step 2 and Step 3):

\begin{enumerate}
    \item First (Section~\ref{subsec:SearchLocalExtrema}), we search for the positions in~${\bf r}_{mn}$ corresponding to local maxima and local minima of the surface $z=\eta({\bf r}, t_0)$; keeping only the set of local maxima with positive $z$, denoted by~${\bf r}^+_{mn}$, and the set of local minima with negative $z$, denoted by~${\bf r}^-_{mn}$.
    \item Second (Section~\ref{subsec:DelaunayTerrain}), we appropriately connect the points~${\bf r}^+_{mn}$ to compute a piecewise-linear, i.e., polyhedral, surface~${\mathcal D}^+$ which approximates the upper wave envelope. Applying the same process to the points~${\bf r}^-_{mn}$, we get an approximation~${\mathcal D}^-$ to the lower wave envelope.
    \item Finally (Section~\ref{subsec:RefinementSplines}), these piecewise-linear surfaces ${\mathcal D}^+$ and ${\mathcal D}^-$ are refined to continuous surfaces by using \emph{tensor product splines}.
\end{enumerate}


\subsection{Search of local extrema}\label{subsec:SearchLocalExtrema}

\begin{figure}[ht]
\begin{center}
\includegraphics[width=0.6\textwidth]{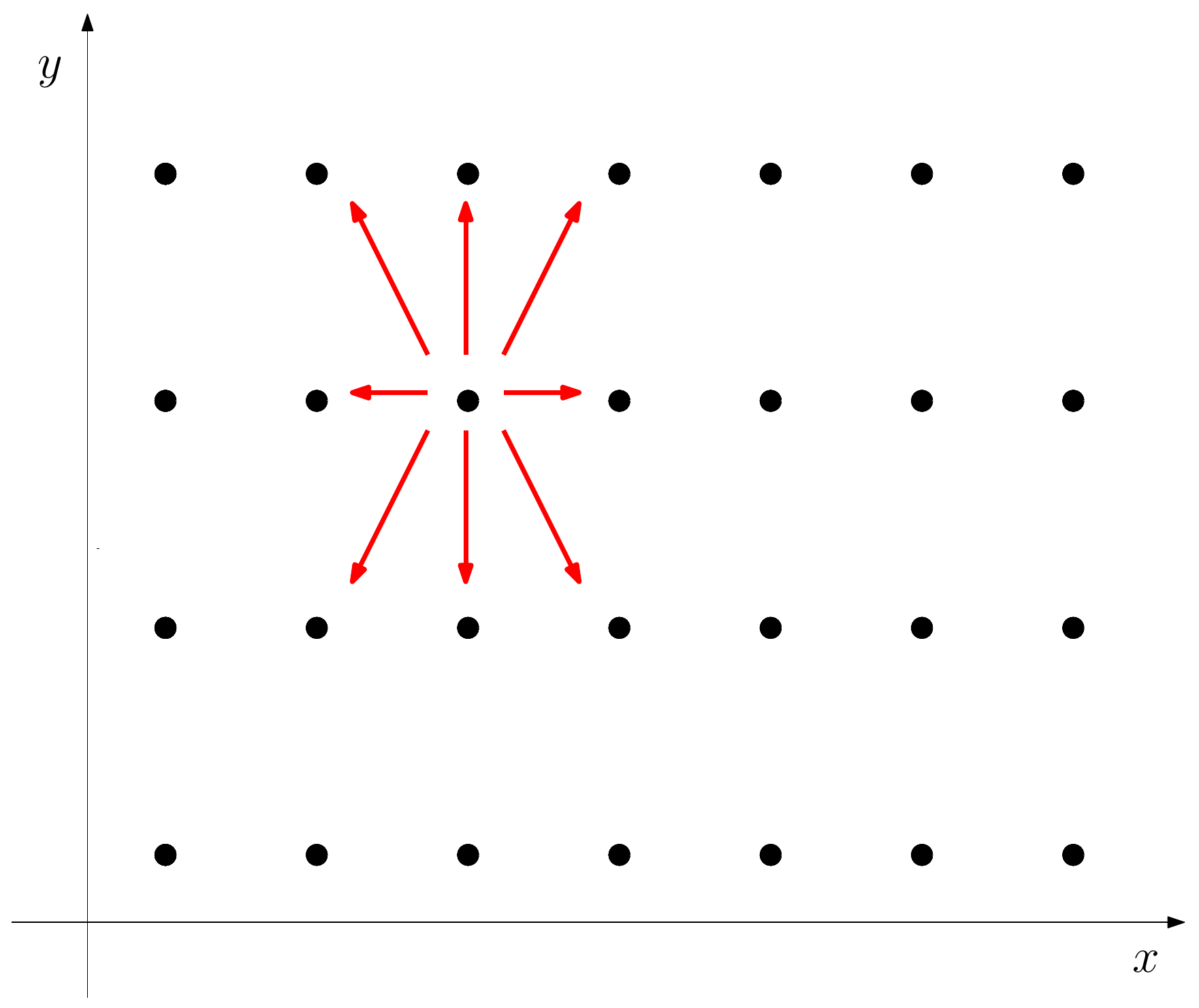}
\end{center}
\caption{Comparing elevation with that of neighbors.\label{ElevationNeighbors}}
\end{figure}

In the first step, our method runs over the points ${\bf r}_{mn}=(x_m,y_n)=(m\Delta x,n\Delta y)$ for $m=0,1,\ldots,N_x-1$ and $n=0,1,\ldots N_y-1$, comparing the elevation~$\eta_{mn}$ of each~${\bf r}_{mn}$ with the elevations of its neighbors in the eight directions N, NW, W, SW, S, SE, E, NE (note that positions on the boundary of the mesh actually have less than eight neighbors); see Fig. \ref{ElevationNeighbors}.

Then we label as local maxima (respectively minima) those positions ${\bf r}_{mn}$ whose elevation is greater (respectively smaller) than the elevations of the neighbors. Finally, we discard those local maxima with negative elevation and those local minima with positive elevation, to obtain the sets~${\bf r}^+_{mn}$ and~${\bf r}^-_{mn}$ of positive local maxima and negative local minima.

\subsection{Delaunay polyhedral terrain}\label{subsec:DelaunayTerrain}

\begin{figure}[htb]
\begin{center}
\subcaptionbox{In red, subset ${\bf r}^+_{mn}$ of the mesh\label{SubsetLocalMaxima}}
{\includegraphics[width=0.45\textwidth]{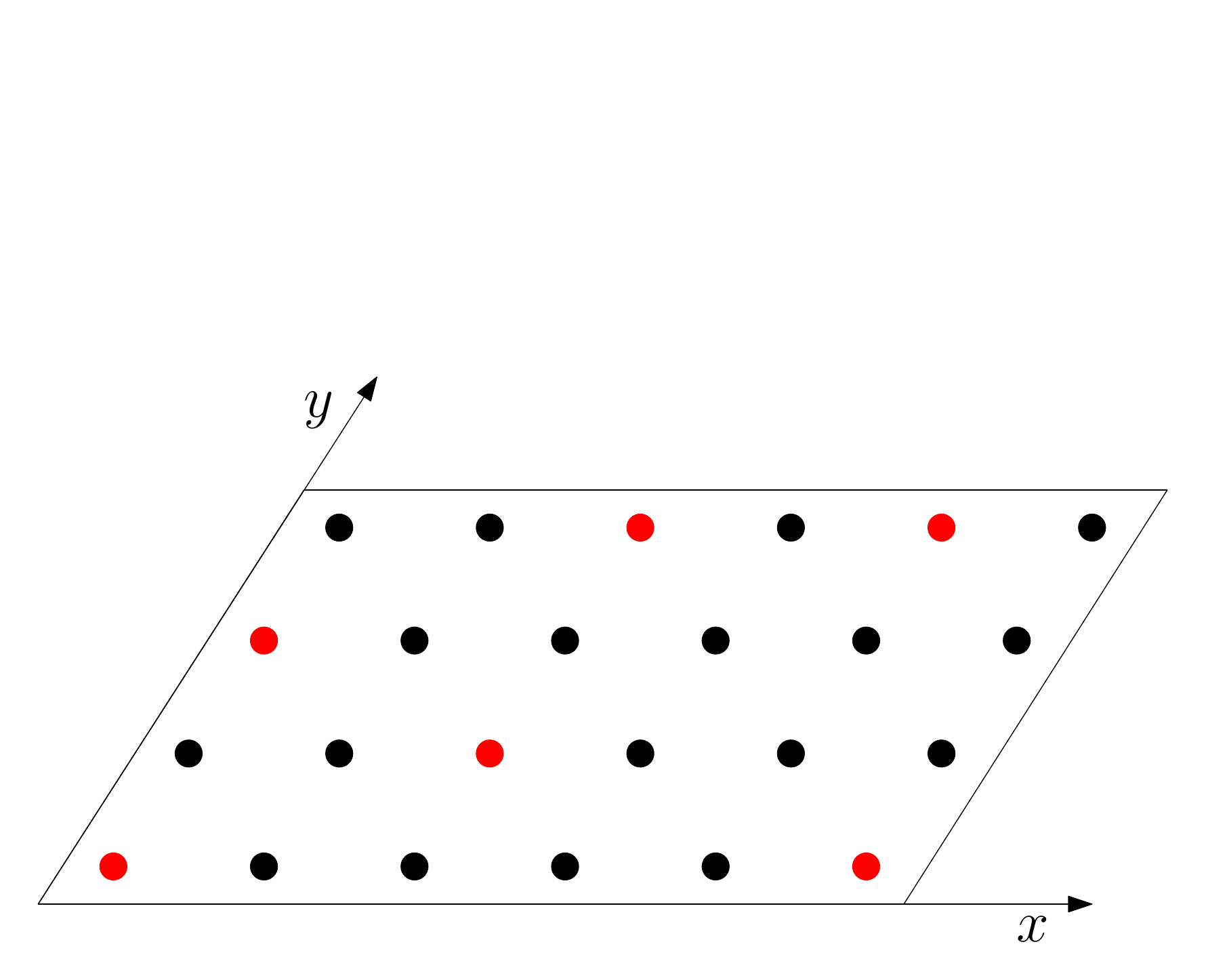}}
\subcaptionbox{Delaunay triangulation of~${\bf r}^+_{mn}$\label{DelaunayTriangulation}}
{\includegraphics[width=0.45\textwidth]{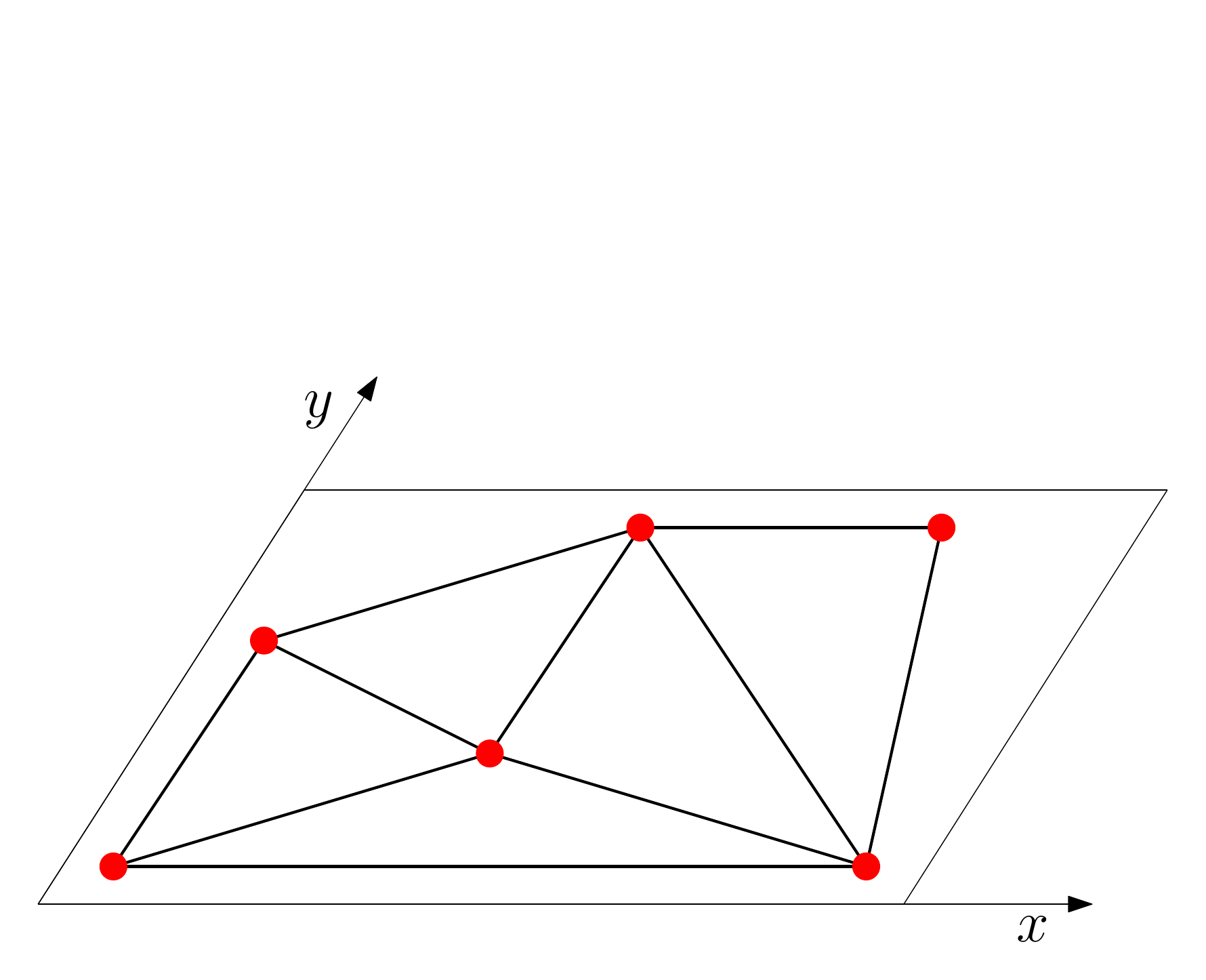}}

\smallskip

\subcaptionbox{Lifting one triangle\label{Lifting1}}
{\includegraphics[width=0.45\textwidth]{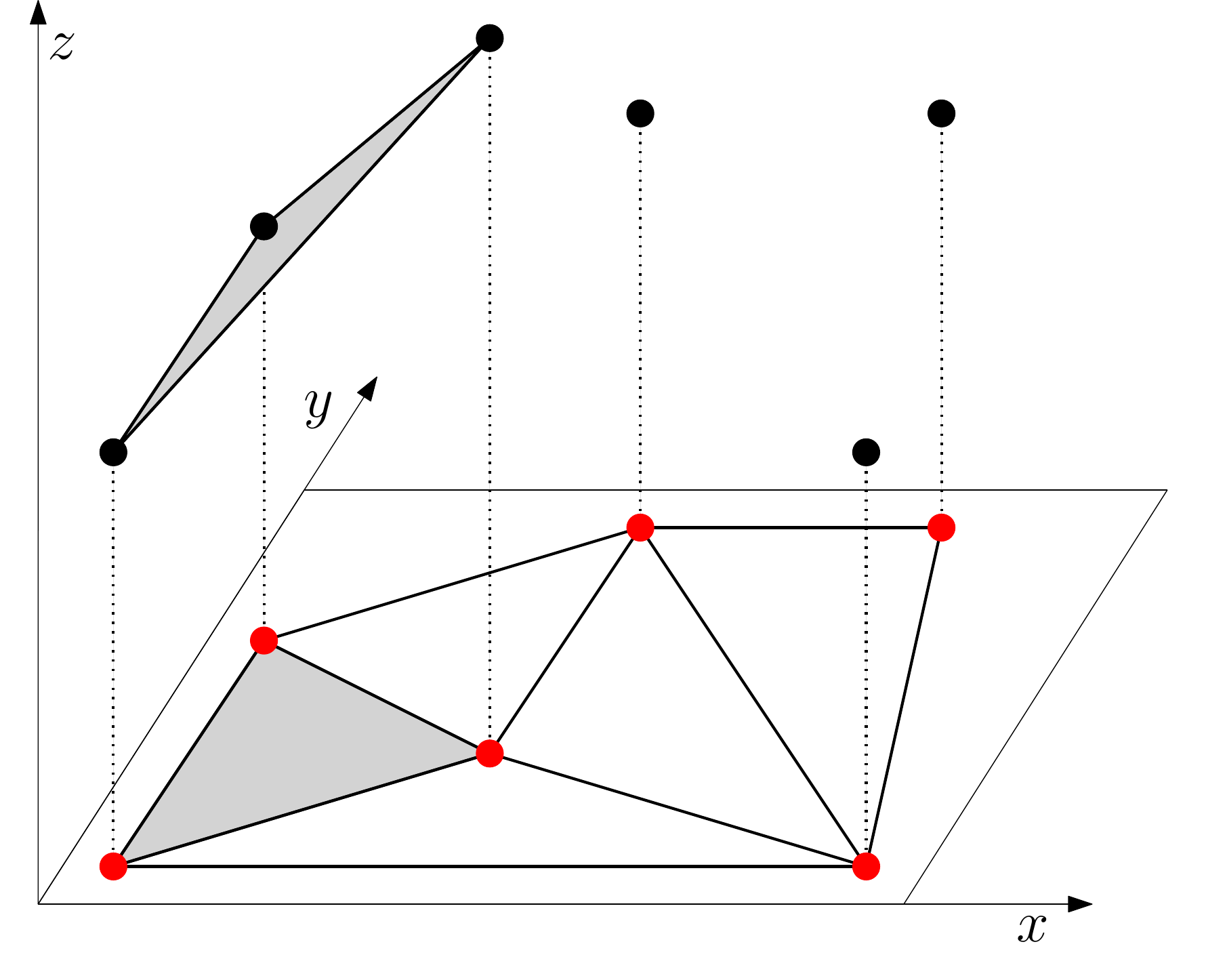}}
\subcaptionbox{Lifted triangulation\label{LiftedTriangulation}}
{\includegraphics[width=0.45\textwidth]{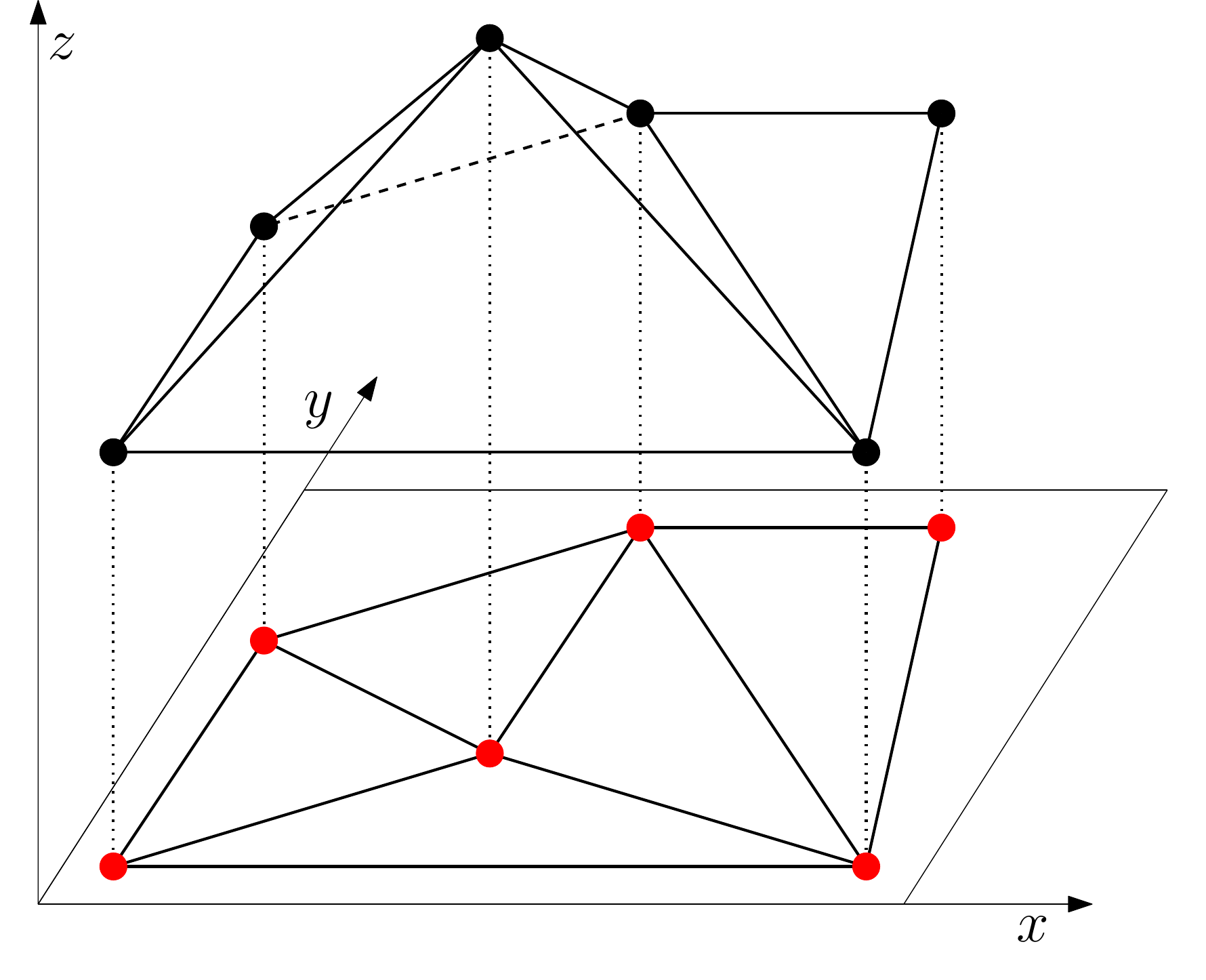}}

\end{center}
\caption{Construction of an envelope as a Delaunay polyhedral terrain.}\label{fig:DelaunayPolyhedralTerrain}
\end{figure}

The second step of our method constructs a first approximation of the envelopes, with the form of a polyhedral terrain, a structure widely used in Geographic Information Systems~\cite{Floriani2000}. This structure interpolates a set of points in~3D, in our case the points~$({\bf r}^+_{mn},\eta_{mn})$ (we will do the same later for~$({\bf r}^-_{mn},\eta_{mn})$), by a piecewise-linear surface composed of triangles whose vertices are those 3D~points. The whole process is illustrated in Fig.~\ref{fig:DelaunayPolyhedralTerrain}.

We first have to triangulate the 2D projections of those points. In our case these are the positive local maxima positions~${\bf r}^+_{mn}$ from the previous step. Although there are many possibilities to triangulate a point set, it has been proved that long and skinny triangles must be avoided for a good interpolation~\cite{Barnhill77}.

\begin{figure}[ht]
\begin{center}
\includegraphics[width=0.7\textwidth]{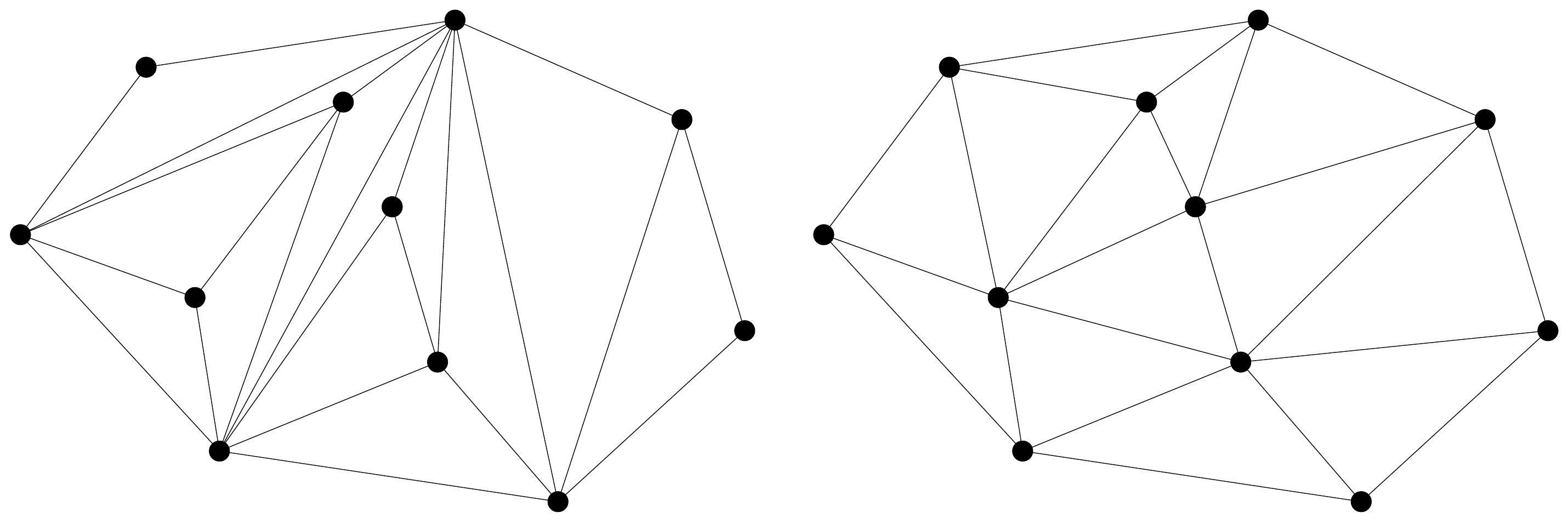}
\end{center}
\caption{Left: Triangulation with long and skinny triangles. Right: Delaunay triangulation of the same point set.\label{DelaunayVsSkinny}}
\end{figure}

A natural way to avoid such undesired triangles is to avoid very small angles. If one looks at the smallest angle of each possible triangulation and chooses the triangulation in which the smallest angle is the maximum (among all the possible triangulations), this is by definition the Delaunay triangulation~\cite{Fortune2004}. See Fig.~\ref{DelaunayVsSkinny}. Being a key tool in Computational Geometry~\cite{deBerg2008}, the Delaunay triangulation has a number of other applications apart from polyhedral terrains.

Our aimed polyhedral terrain will be then the result of lifting this 2D Delaunay triangulation of~${\bf r}^+_{mn}$ to 3D. The triangles in 2D are lifted to 3D by just lifting the points~${\bf r}^+_{mn}$ to~$({\bf r}^+_{mn},\eta_{mn})$. See Fig.~\ref{Lifting1}. This leads to a series of two-dimensional triangles tilted in 3D, see Fig.~\ref{LiftedTriangulation}, which constitutes our approximation of the upper wave envelope~${\mathcal D}^+$.

Applying the same process to the negative local minima positions~${\bf r}^-_{mn}$, we obtain an approximation~${\mathcal D}^-$ of the lower wave envelope.

\begin{figure}[htb]
  \centering
 \includegraphics[width=\hsize]{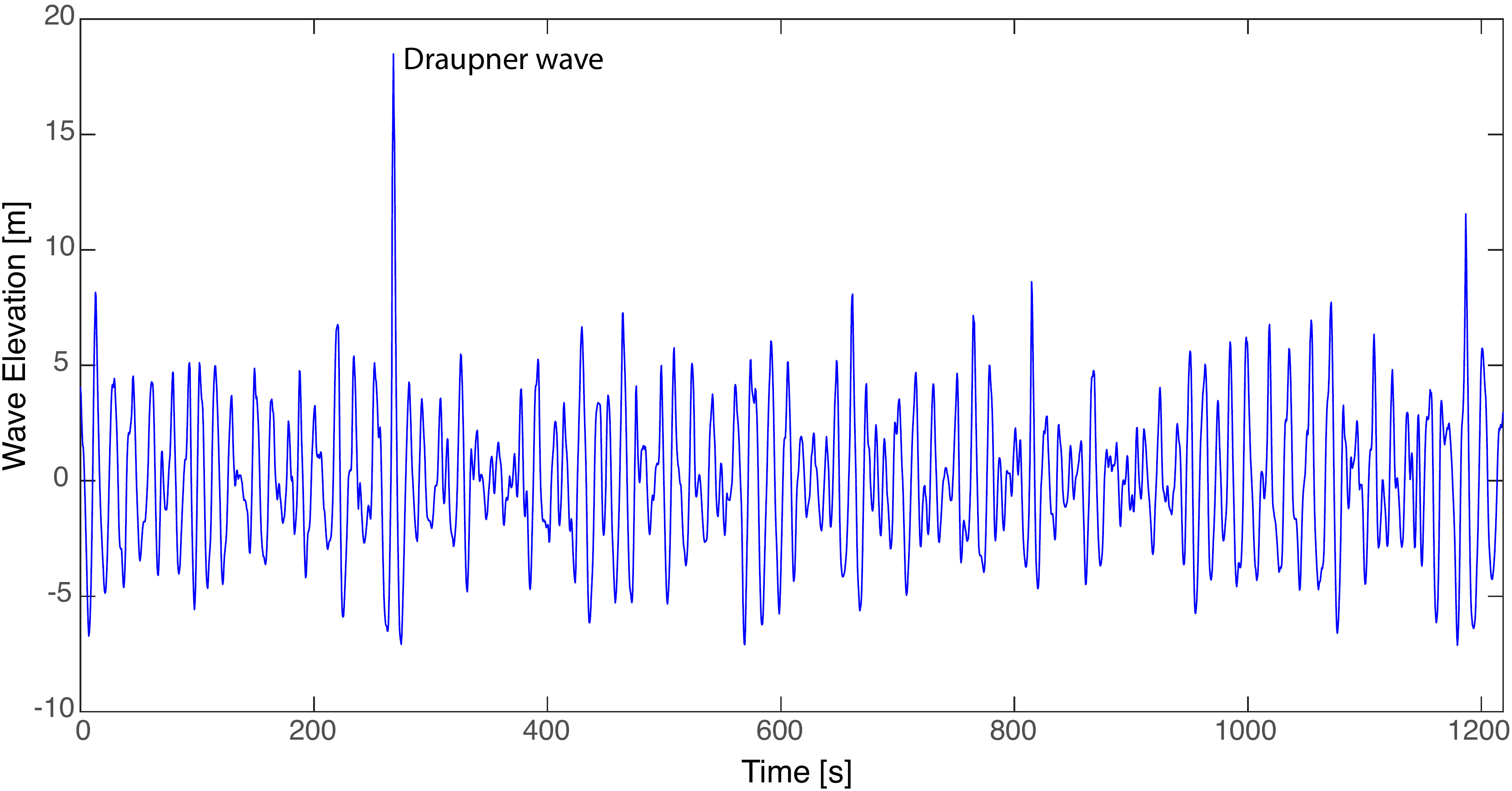}
  \caption{Draupner wave record measured at to the Draupner platform on 1 January 1995 in the North Sea \cite{Hansteen2003}.}\label{fig:DraupnerRecord}
\end{figure}

\begin{figure}[htb]
  \centering
 \includegraphics[width=\hsize]{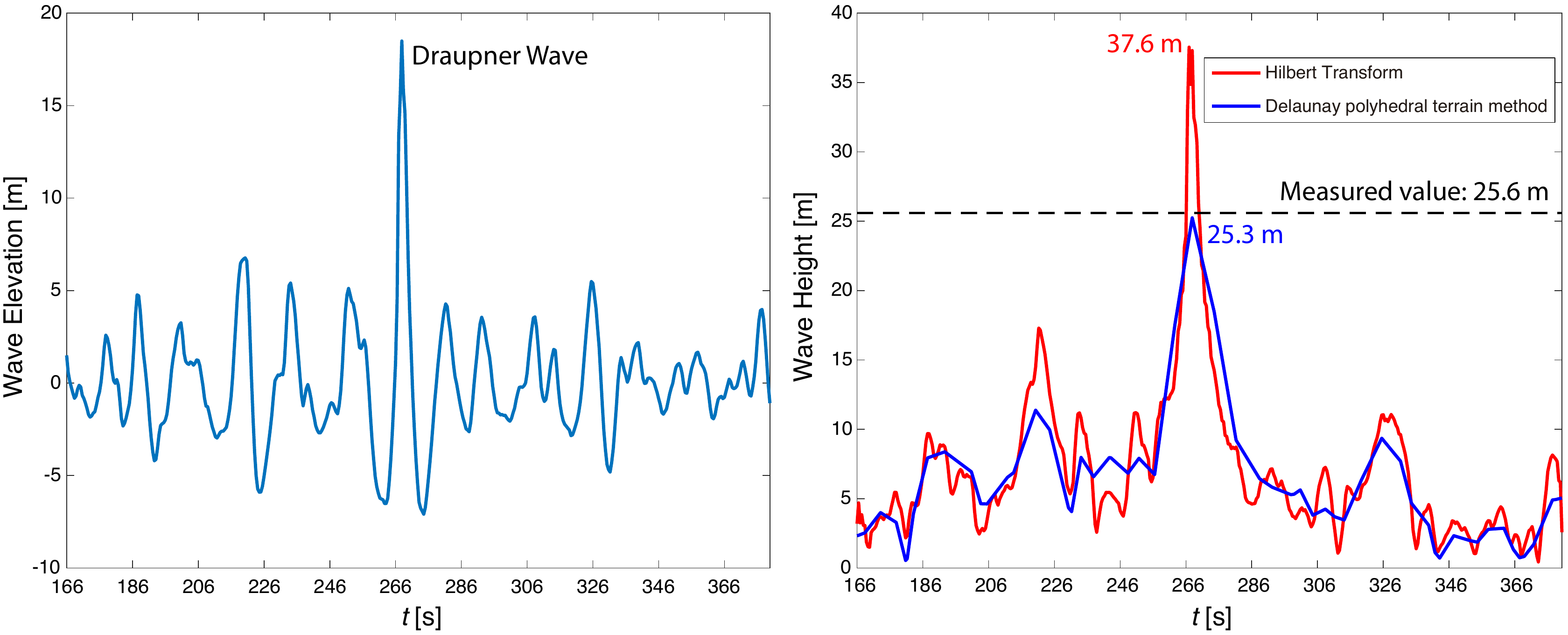}
  \caption{Subset of the Draupner wave record (left) and the corresponding estimation of the local wave height (right) from the Hilbert transform (red line) and the Delaunay polyhedral terrain method (blue line). The dashed line indicates the published value of the height of the freak wave \cite{Hansteen2003,Walker2004}.}\label{fig:DraupnerWave}
\end{figure}

To illustrate the advantages of the application
of the Delaunay polyhedral terrain method to the estimation of the local wave height,
the method has been used to the particular case of a one-dimensional data set (i.e. a time series record of wave elevations $\eta(t)$).
In the same way, the estimation of the significant wave height derived from the Riesz transform or, more precisely for one dimension, the Hilbert transform, has been estimated as well.
The used data set corresponds to the well-known historical Draupner wave record
measured on 1 January 1995 in the Draupner oil platform ~\cite{Hansteen2003}.
The Draupner platform is located in the Norwegian sector of the North Sea in a water depth of 70~m.
This measurement registered a freak wave of 25.6~m for a sea state with a significant wave height of approximately 12~m.
Figure~\ref{fig:DraupnerWave} shows the wave elevation time series of this record, where the freak wave, commonly known in the literature as Draupner wave or New Year Wave~\cite{Walker2004} is labeled.
Figure~\ref{fig:DraupnerWave} (left) shows a subset of record for times around the freak wave event.
The estimation of the local wave height by the Hilbert transform  and the corresponding estimation using the  Delaunay polyhedral terrain approach are shown in the right part of Figure~\ref{fig:DraupnerWave}.
Since the wave record does not present statistical symmetry between wave crests and throughs (see Figure~\ref{fig:DraupnerRecord}), it is not a good approach to estimate the local wave height using the envelope estimation from the Hilbert Transform.
Hence, the estimation of the local wave height (i.e. 37.6~m) from the Hilbert transform is higher than height measured in Draupner, whereas the estimation derived from the Delaunay polyhedral terrain method (i.e. 25.3~m) is closer to the value derived from standard techniques applied to wave elevation time series.

\subsection{Refinement by splines}\label{subsec:RefinementSplines}

\begin{figure}[htb]
  \centering
  \[\begin{array}{ccc}
 \includegraphics[width=4cm,height=4cm]{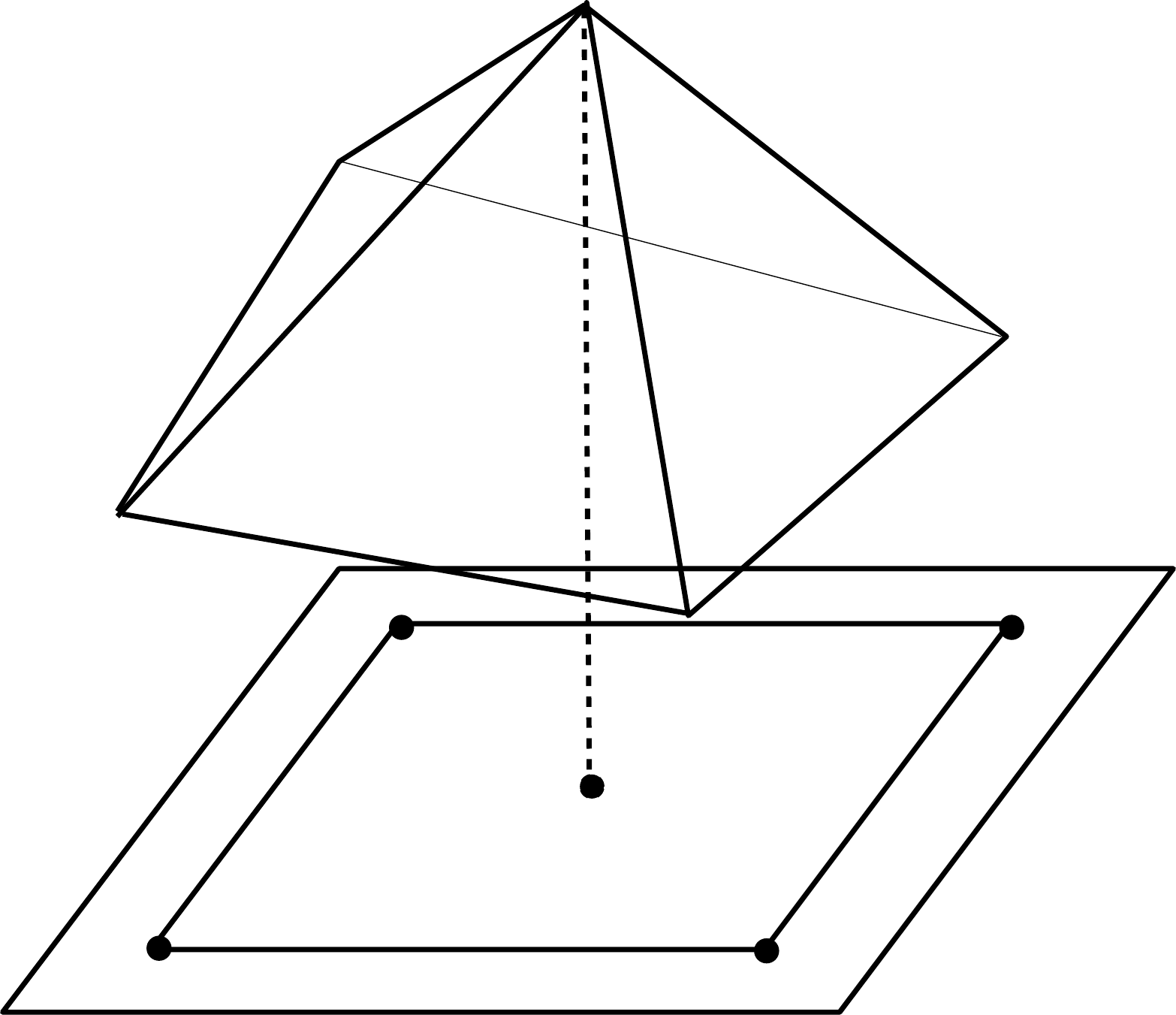} &\includegraphics[width=4cm,height=4cm]{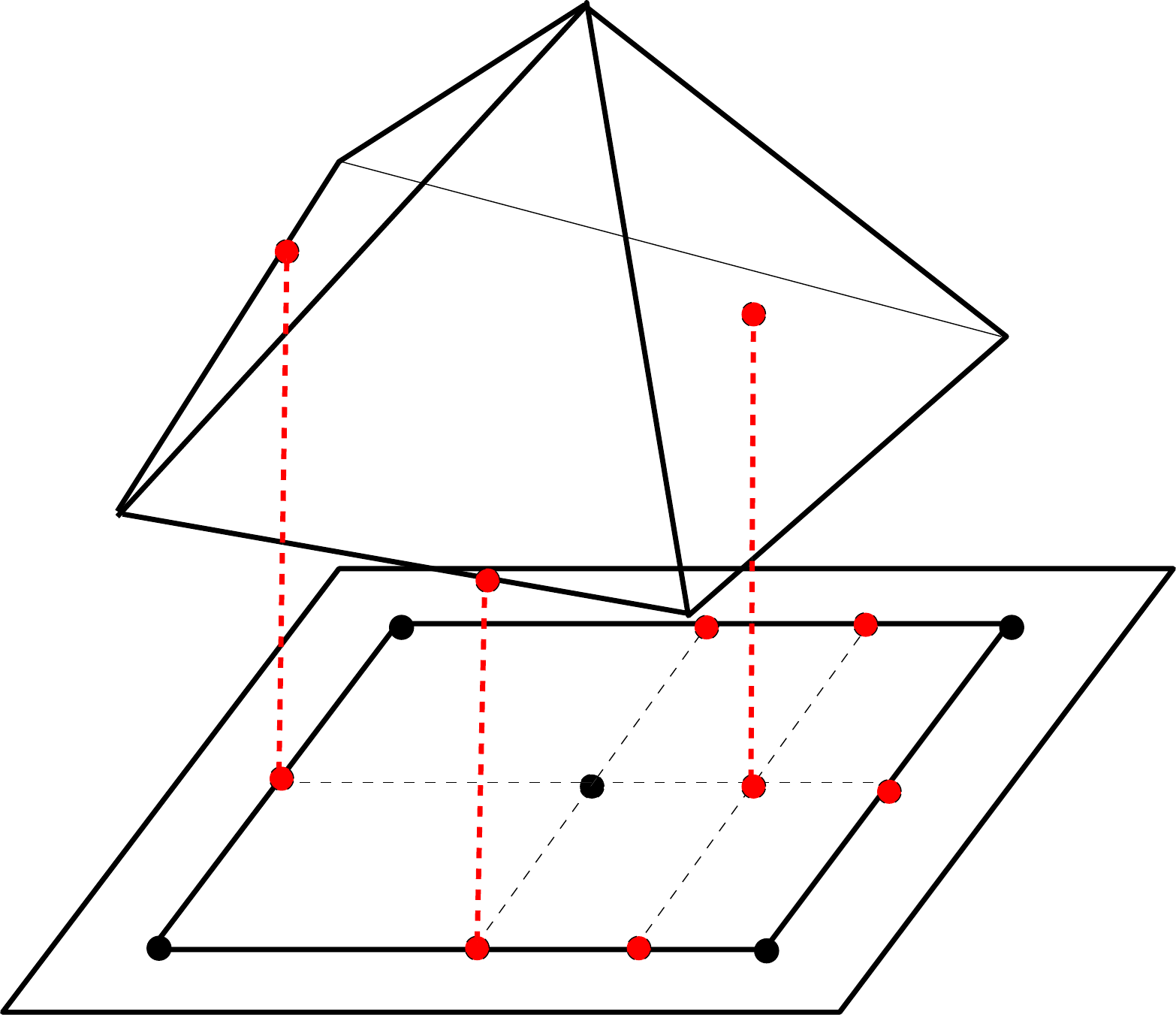} & \includegraphics[width=4cm,height=4cm]{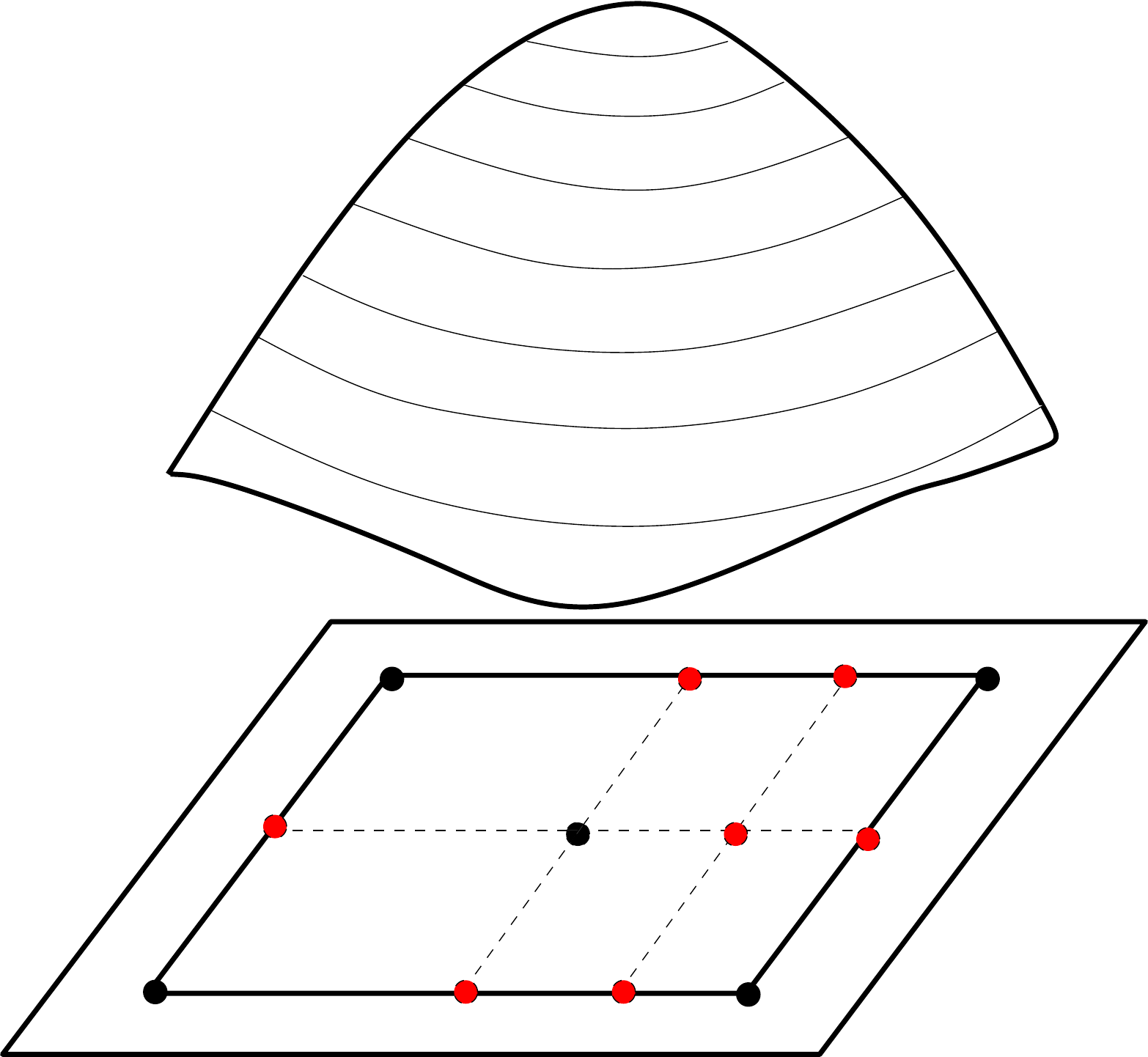}
  \end{array}\]
  \caption{Refinement with Splines.}\label{fig:splines}
  \end{figure}

The third step of our method refines the piecewise-linear surfaces~${\mathcal D}^+$ and~${\mathcal D}^-$ computed in the previous step, by using spline methods (see Fig. \ref{fig:splines}).
To do this, we use \emph{tensor product spline surfaces} \cite{Lyche-Morken08}. Essentially, a tensor product spline surface is an explicit surface $z=f(x,y)$, where
\begin{equation}\label{tensor}
f(x,y)=\sum_{p=1}^{m_1} \sum_{q=1}^{m_2} c_{pq}\phi_p(x) \psi_q(y).
\end{equation}
The functions $\phi_p(x)$, $\psi_q(y)$ are \emph{$B$-spline basis functions} \cite{Lyche-Morken08,Patrik02}. These are piecewise polynomial functions, widely used in Computer Aided Geometric Design and Interpolation Theory, that are defined in a recursive way. In our case we have used cubic splines, i.e., each $\phi_p(x)$ and $\psi_q(y)$ have degree 3.

Given a set of points $\tilde{p}_{mn}=(\tilde{x}_m,\tilde{y}_n)$ forming a rectangular grid (as in Fig. \ref{ElevationNeighbors}), and the values $\tilde{z}_{mn}$ of a variable $z$, measured at the points $\tilde{p}_{mn}$, one can determine a tensor product spline surface interpolating the space points $\tilde{P}_{mn}=(\tilde{x}_m,\tilde{y}_n,\tilde{z}_{mn})$ \cite{Lyche-Morken08}. To do this, one computes the coefficients $c_{pq}$ in \eqref{tensor} by solving the matrix equation \cite[Proposition~7.3]{Lyche-Morken08}
\[{\bf \Phi}\cdot {\bf C}\cdot {\bf \Psi}^T={\bf F},\]
where ${\bf C}=(c_{pq})$, ${\bf \Phi}=(\phi_{mp})$, with $\phi_{mp}=\phi_p(x_m)$, ${\bf \Psi}=(\psi_{nq})$, with $\psi_{nq}=\psi_q(y_n)$, and ${\bf F}=(\tilde{z}_{mn})$. Since the elements of ${\bf \Phi}$, ${\bf \Psi}$, ${\bf F}$ are known, one has

\begin{equation}\label{splines-eq}
{\bf C}={\bf \Phi}^{-1}\cdot {\bf F} \cdot {\bf \Psi}^{-T}.
\end{equation}

In our case, we take $\tilde{p}_{mn}={\bf r}_{mn}$, i.e., we use the planar rectangular grid of Fig. \ref{ElevationNeighbors}. Now we want to refine the polyhedral surfaces ${\mathcal D}^+$ and ${\mathcal D}^-$ constructed in the previous subsection, to get two new piecewise-polynomial surfaces ${\mathcal S}^+$ and ${\mathcal S}^-$, of degree 3 in $x$ and $y$. In order to compute ${\mathcal S}^+$ (the computation of ${\mathcal S}^-$ is analogous), we proceed in the following way: If $\tilde{p}_{mn}\in {\bf r}_{mn}^+$, we take $\tilde{z}_{mn}=\eta_{mn}$; if $\tilde{p}_{mn}\notin {\bf r}_{mn}^+$, we take $\tilde{z}_{mn}$ as the $z$-value corresponding to $\tilde{p}_{mn}$ in~${\mathcal D}^+$. Then ${\mathcal S}^+$ is the tensor product spline surface \eqref{tensor} that interpolates the points $\tilde{P}_{mn}=({\bf r}_{mn},\tilde{z}_{mn})$.

\subsection{Complexity of our methods}

Let~$N=N_xN_y$ be the number of points in the original mesh~${\bf r}_{mn}$. The computational complexity of our discrete method computing~${\mathcal D}^+$ and~${\mathcal D}^-$ is in~$O(N\log N)$: Searching for local extrema is~$O(N)$, computing the Delaunay triangulation is~$\Theta(N\log N)$~\cite{deBerg2008} and the lifting process is again~$O(N)$.

The computational complexity of our spline method constructing ${\mathcal S}^+$ and~${\mathcal S}^-$ is in~$O(N^{3/2})$: This is the complexity for both multiplication and computation of the inverse for square matrices with~${\mathcal N}$ entries and, hence, the complexity of computing the matrix ${\bf C}$. Since this last complexity dominates the other ones, we get an overall complexity for our method of $O(N^{3/2})$.

\subsection{Additional observations on the grid.} \label{add-obs}

Although for simplicity in the previous subsections we considered a rectangular grid, our method can be used for other grids as well. The only property needed is that, for each point, its set of neighbors is clearly defined. Then, we can compare the elevation at each point with the elevations of its neighbors, in order to find local maxima and minima. For example, the reader can imagine the analogous to Figure~\ref{ElevationNeighbors} with a triangular grid or an hexagonal grid (note that a tiling is a sufficient, not necessary, condition for the set of neighbors to be determined). Additionally, notice that we do not need the grid to be regular or equiespaced in any direction, whenever the notion of ``set of neighbors'' is clearly determined for each point.

\section{Application of the method}
\label{sc:Experiments}

\subsection{Description of the used wave field simulation techniques}
\label{sc:Simulations}
To estimate the envelope by using the discrete and spline methods described above, a weak non-linear method is based on the stochastic description of ocean waves have been used.
The simulation method consists on two steps:
The first one consists of the simulation of a linear Gaussian wave field based in the model described by \eqref{eq:SeaState}.
The second step takes into account weak linear second order contribution to the wave field \cite{Tayfun80}.
The two steps are described on the following:

\begin{itemize}
	\item[-] Linear wave fields: it is a very well-known technique based on \eqref{eq:SeaState}. Thus, the simulated linear wave field $\eta_l$ is regarded as
	
	\begin{equation}\label{eq:SeaStateSimLin}
		\eta_l({\bf r}, t) = \sum_{{\bf k}} a_{\bf k} \cos \left( {\bf k} \cdot {\bf r} - \omega({\bf k}) t + \varphi_{\bf k} \right)
		\mbox{,}
	\end{equation}
	
    \noindent
    where
    $\omega({\bf k})$ is the dispersion relation for linear gravity waves.
    In this paper deep water conditions have been considered, $\omega({\bf k}) = \sqrt{g k}$. The phases~$\varphi_{\bf k}$ in \eqref{eq:SeaStateSimLin} are uniform distributed in $[-\pi, \pi)$. The amplitudes~$a_{\bf k}$ are derived from the wave number spectrum $F({\bf k})$ as

	\begin{equation}\label{eq:SimAmplitudes}
		a_{\bf k} = \sqrt{F({\bf k}) \, \Delta k_x \Delta k_y  \cdot \left( \alpha_{\bf k}^2 + \beta_{\bf k}^2 \right) \,}
		\mbox{,}
	\end{equation}

    \noindent where $\alpha_{\bf k}$, and $\beta_{\bf k}$ are uncorrelated zero-mean Gaussian processes of variance 1 \cite{Rodriguez04,Nieto-Borge05}. Under these conditions, the amplitudes $a_{\bf k}$ are Rayleigh distributed. $\Delta k_x$ and~$\Delta k_y$ are the wave number resolutions for each ${\bf k}$-axes. Assuming that Eq. \ref{eq:SeaStateSimLin} is computed by using a two-dimensional FFT algorithm for each time $t$, $\Delta k_x$ and $\Delta k_y$
        are given by the required spatial resolutions of the simulated linear wave field, $\Delta x$, $\Delta y$, and the number of samples ($N_x$ and $N_y$) of the simulated sea surface of total area, $N_x \Delta x \cdot N_y \Delta y$.
    In practice, the spectrum $F({\bf k})$ in Eq. \eqref{eq:SimAmplitudes} is derived from the frequency spectrum $S(\omega)$ and the directional spreading function $D(\omega, \theta)$ by using the dispersion relation $\omega({\bf k})$,

    \begin{equation}\label{eq:FfromSandD}
    	F({\bf k}) = S(\omega) D(\omega, \theta) k^{-1} c_g({\bf k})
    	\mbox{,}
    \end{equation}

    \noindent
    where $\theta = \arg {\bf k}$ is the wave propagation direction.
    Eq. \eqref{eq:FfromSandD} considers the appropriate Jacobians needed for the change of coordinates $(\omega, \theta) \mapsto (k_x, k_y)$ to keep the total variance of the stochastic process $\eta_l$.
    Hence, $c_g({\bf k}) = d \omega ({\bf k}) / d k$ is the group velocity, which is the Jacobian needed for the change of coordinates $\omega \mapsto k$, and
        $k^{-1}$ is the Jacobian needed for the change of polar $(k, \theta)$ to the wave number vector Carthesian coordinates $(k_x, k_y)$.
        Under the conditions described above, the model given by \eqref{eq:SeaStateSimLin} is Gaussian distributed.

	\item[-] Weak non-linear wave fields: For weak nonlinear wave fields, it is possible to take into account Stokes wave expansions, where there are high-order interactions between different wave components \cite{Ochi2005}.
	Therefore, the week nonlinear wave field based on second order Stokes waves $\eta_{nl}$ is given by

	\begin{equation}\label{eq:NonLinExpansion}
		\eta_{nl}({\bf r}, t) = \eta^{(1)}({\bf r}, t) + \eta^{(2)}({\bf r}, t) + \ldots
		\mbox{,}
	\end{equation}

    \noindent
    where the upper index indicates the order of the expansion term. The first order term corresponds to the linear model given by \eqref{eq:SeaStateSimLin}, i.e. $\eta^{(1)} = \eta_l$.
    The second order term $\eta^{(2)}({\bf r},t)$ contains the contributions of the summation $({\bf k} + {\bf k}')$ and difference $({\bf k} - {\bf k}')$ interactions between different wave components \cite{Tayfun+Fedele2007,Fedeleetal2010,Fedeleetal2011a}.
    The simulation of wide banded process of the second order contribution $\eta^{(2)}$ requires the use of quadratic transfer functions for the interaction between all the different wave number components \cite{Hasselmann1962,JhaWinterstein2000}. In the two-dimensional case, the use of those quadratic transfer functions needs very high CPU requirements, because it is necessary to compute two two-dimensional FFTs for each wave number component~${\bf k}$ to estimate all the $({\bf k} + {\bf k}')$ and $({\bf k} - {\bf k}')$ possible interactions \cite{Nieto-Borge05}. A technique to reduce the CPU time consists of assuming that the nonlinear wave field is a narrow-banded process. This method was originally developed by \cite{Tayfun80} to determine the wave height probability density function of wave elevation time series. For directional sea states, the narrow band approach of the second order term is given by

        \begin{equation}\label{eq:Tayfun2D}
        \eta^{(2)}({\bf r}, t) =
        \frac{\overline{k}}{2}
        \left[
        \left( \sum_{{\bf k}} a_{\bf k} \cos \Phi_{\bf k} \right)^2 -
        \left( \sum_{{\bf k}} a_{\bf k} \sin \Phi_{\bf k} \right)^2
        \right]
        \mbox{,}
        \end{equation}

      \noindent
      where $\Phi_{\bf k} \myeq {\bf k} \cdot {\bf r} - \omega({\bf k}) t + \varphi_{\bf k}$, and $\overline{k}$ is the mean wave number derived from the wave spectrum $F({\bf k})$.

     Fig.~\ref{fig:HeighLinearNonLinearDelaunay} shows an example of the local have height derived from the Delaunay polyhedral terrain method applied to a linear wave field and its corresponding weak non-linear version.
     The simulation was carried out considering a JONWSAP frequency spectrum with significant wave height of
     $H_s = 2$~m,
     peak angular frequency $\omega_p = 0.2 \pi$~rad/s, peakedness factor $\gamma = 3.3$, $\sigma_a = 0.07$, and $\sigma_b = 0.09$.
     The corresponding directional spreading function $D(\omega, \theta)$ considers the parameterization proposed by \cite{Mitsuyasuetal1975}, where the directional spreading is given by the parameter $s(\omega) = s_{max} (\omega/ \omega_p)^\mu$,
     where $\mu = \mu_1$ for $\omega\le \omega_p$ and $\mu = \mu_2$ for $\omega > \omega_p$.
     In this simulation, the values of the directional spreading parameters are $s_{max} = 15$, $\mu_1 = 5$, and $\mu_2 = -2.5$.
 	
 	One can see that the local wave height estimation of the nonlinear wave field presents local variations compared with the corresponding estimation for the linear waves.

\begin{figure}[htb]
  \centering
 \includegraphics[width=0.8\hsize]{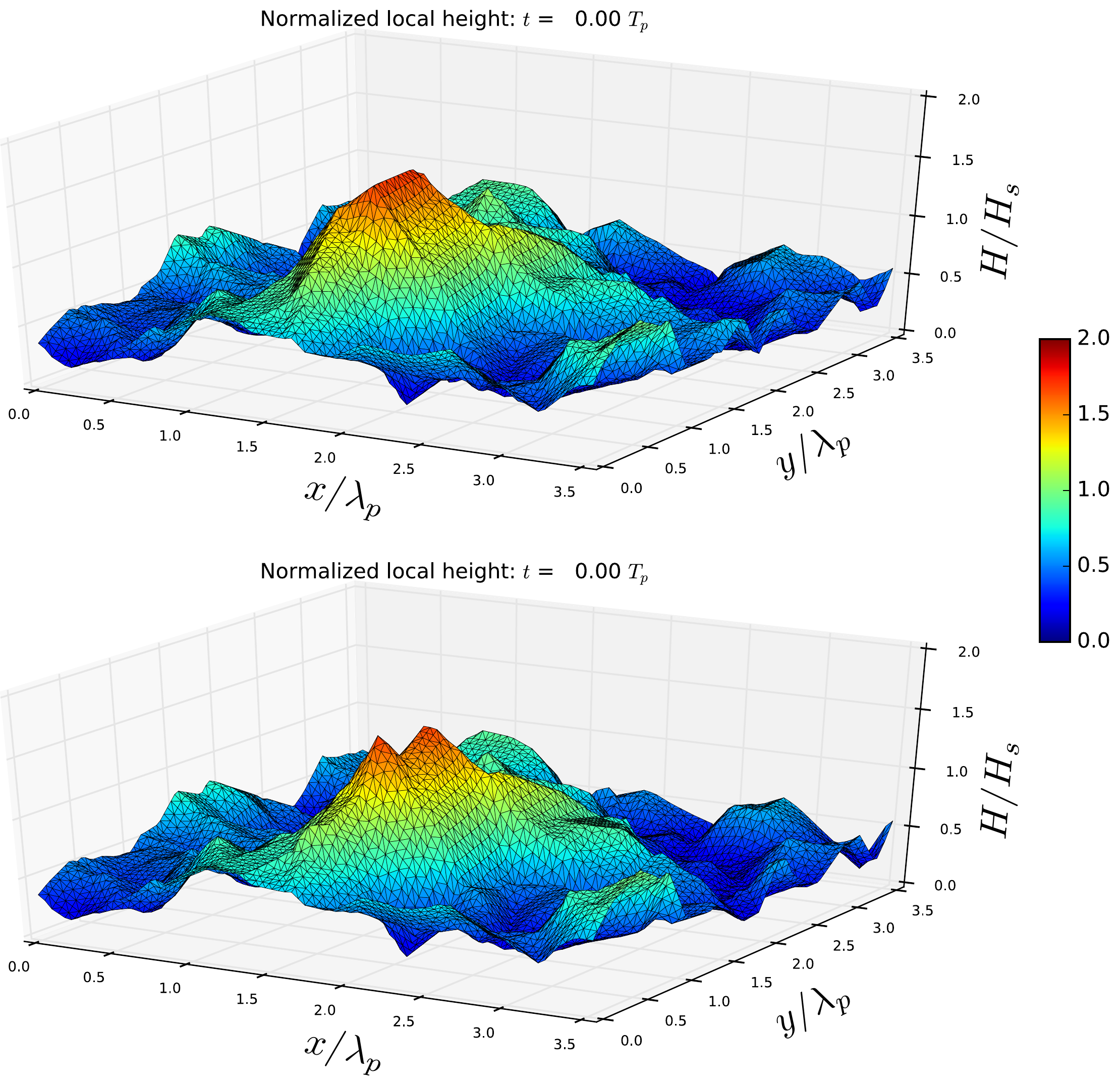}
  \caption{Local heights estimated from the Delaunay polyhedral terrain method applied to a linear wave field (top) and its corresponding non-linear narrow-banded version (bottom). The local height is normalized dividing by the significant wave height $H_s$, and the $x$, $y$-axes by the peak wave length $\lambda_p$.}\label{fig:HeighLinearNonLinearDelaunay}
  \end{figure}

\end{itemize}

\subsection{Obtained results}
\label{sc:Results}

\begin{figure}[htb]
 \includegraphics[width=\hsize]{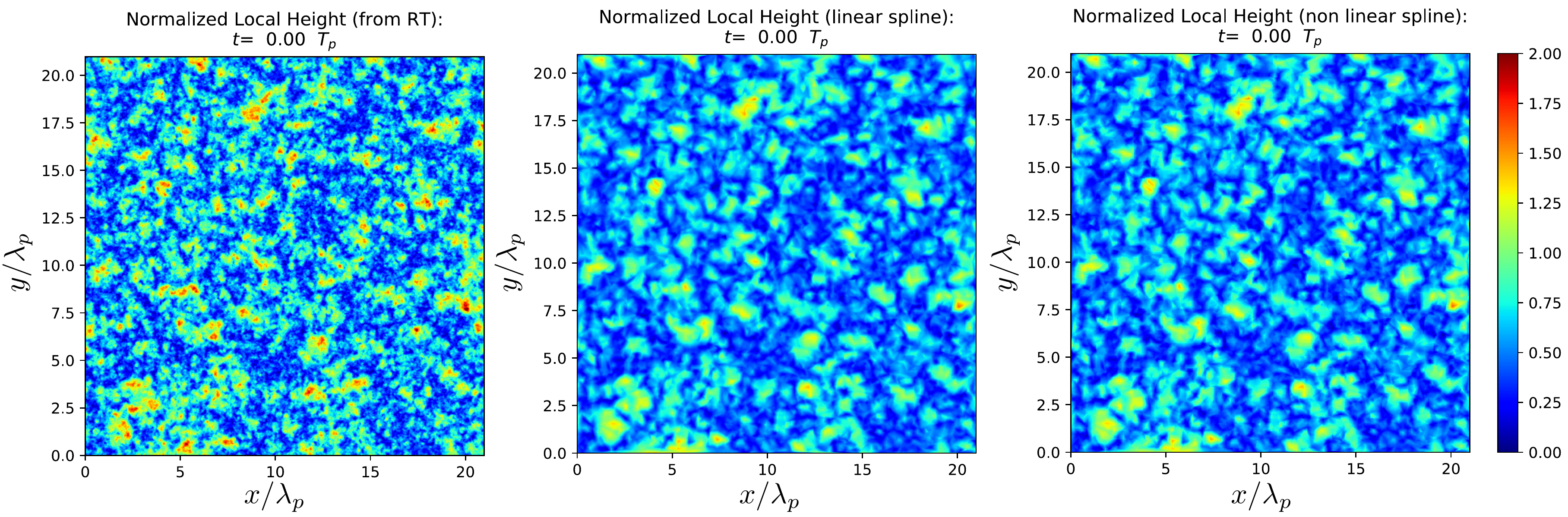}
  \caption{Local height estimated from RT (left), D (center), and S (right).}\label{fig:EnvelopeFields2D}
\end{figure}

\begin{figure}[htb]
  \centering
 \includegraphics[width=0.8\hsize]{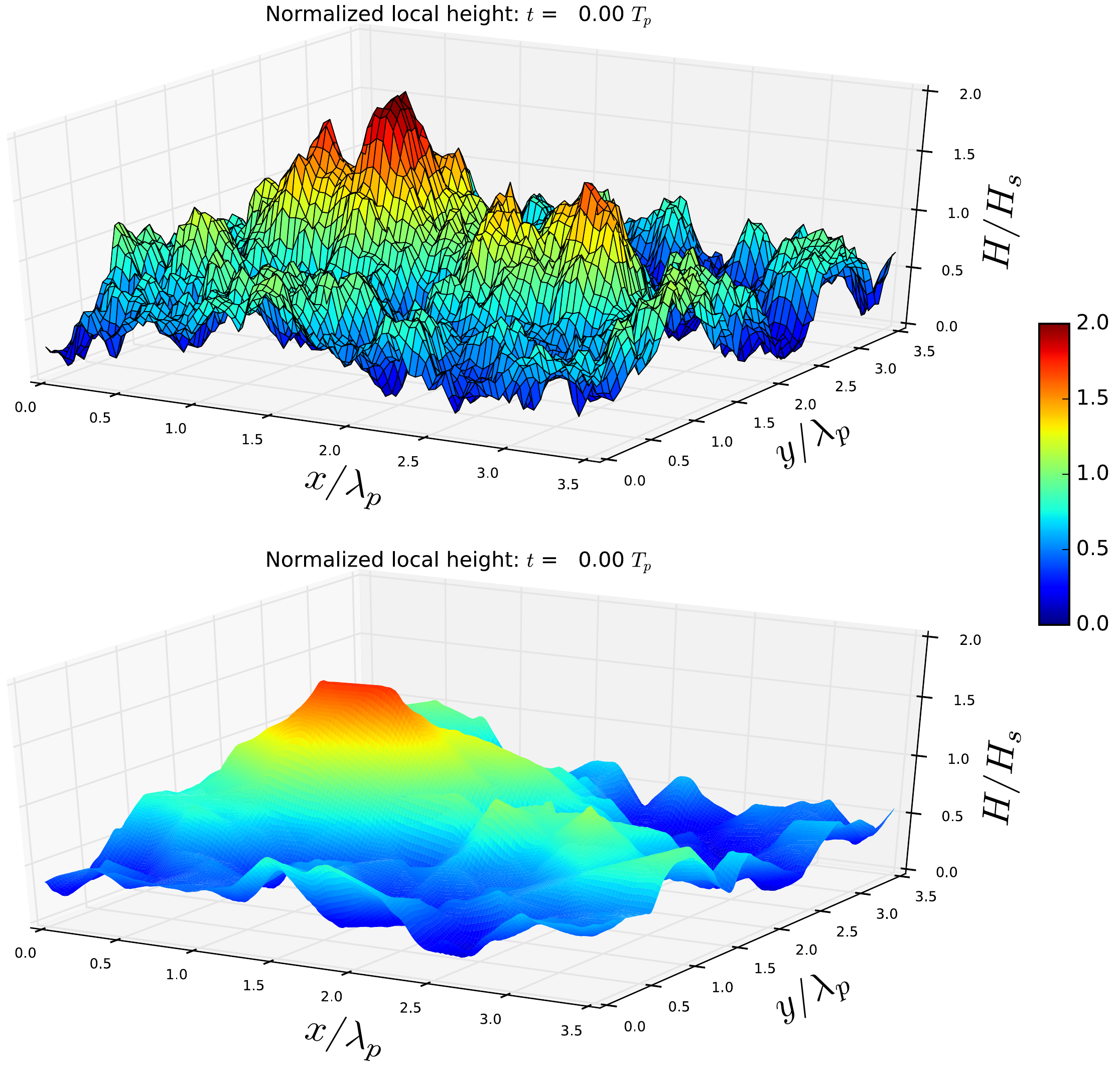}
  \caption{Local wave height estimated from RT (top) and its corresponding estimations from S (bottom).}\label{fig:HeighRTSpline}
  \end{figure}

\begin{figure}[htb]
 \includegraphics[width=\hsize]{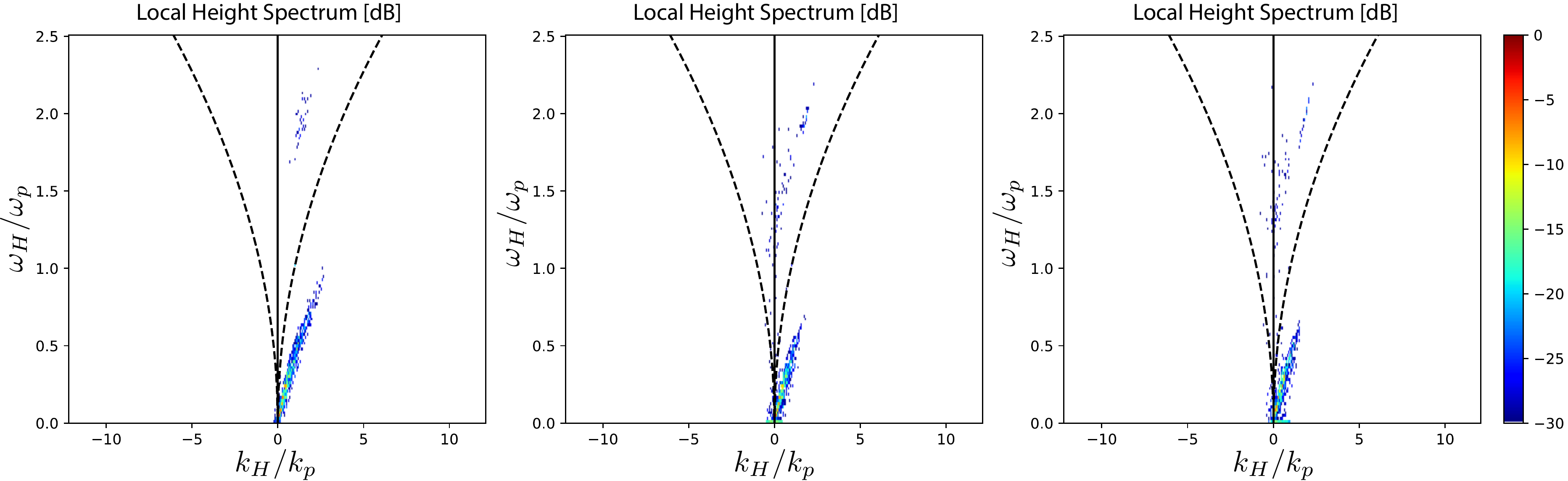}
  \caption{Wave number and frequency spectra from RT (left), D (center), and S (right) in logarithmic scale. The plots are a two-dimensional transect along the wave propagation direction in the three-dimensional space for wave numbers and frequencies. The dashed lines indicates the dispersion relation of linear gravity waves.}\label{fig:EnvelopeSpectrakHw}
\end{figure}

\begin{figure}[htb]
  \includegraphics[width=\hsize]{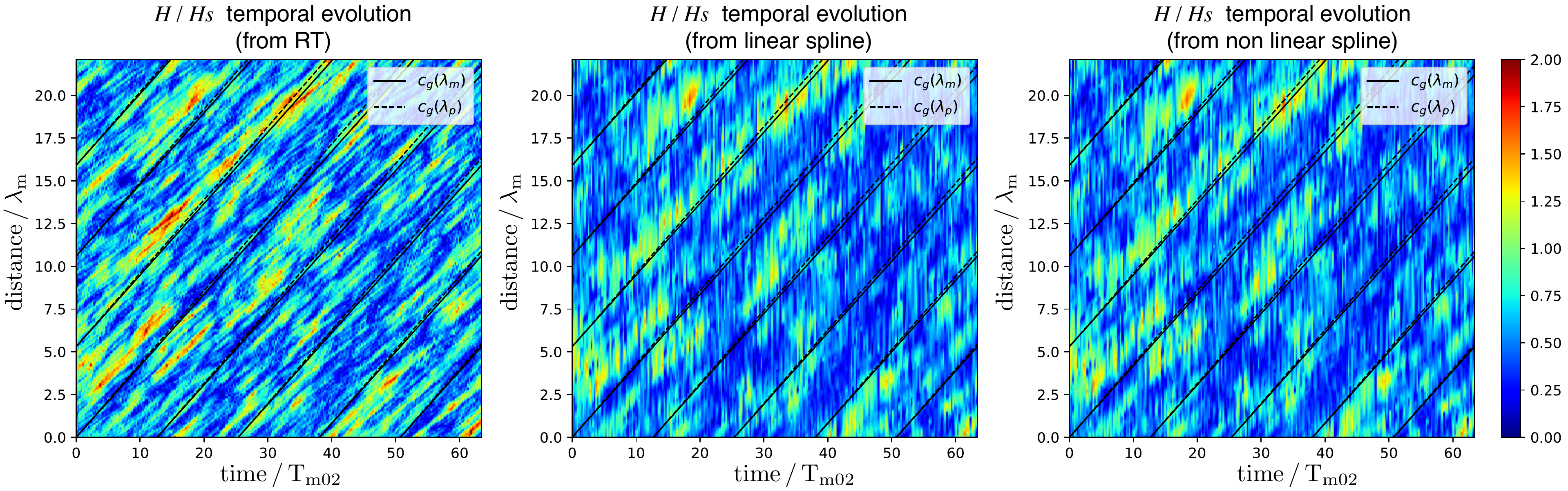}
  \caption{Spatio-temporal evolution of the local height along the main wave propagation direction derived from the two methods: RT (left), D (center) and S (right). The continuous straight lines indicates the group velocity of linear gravity waves for the mean wave length $\lambda_m$. The straight lines correspond to the group velocity for the peak wave length $\lambda_p$. The temporal axis is normalized with the mean period derived from the wave spectrum $T_{m02}$ and the axes of wave propagation distances is normalized with $\lambda_m$.}\label{fig:TemporalEvolution}
\end{figure}

Considering the simulation techniques described in the previous Section~\ref{sc:Simulations}, weak nonlinear $\eta_{nl}$ simulated wave fields have been derived.
For the estimation of the local height $H_{RT}$ derived from RT, the upper and lower envelopes are symmetric (i.e. $A^{+} = - A^{-} = A$). Then, $H_{RT} \sim 2 A$, whereas the estimations for the Delaunay polyhedral terrain (D) and spline (S) based methods (denoted as $F_D$ and $H_S$ respectively), as they can derive asymmetric upper and lower envelopes (i.e. $A^{+}$, $A^{-}$), the local wave height is regarded as $H_{D, S} \sim A_{D, S}^{+} - A_{D, S}^{-}$.
Fig.~\ref{fig:EnvelopeFields2D} shows the local height derived from a weak non-linear wave field using a JONSWAP spectrum for the three methods: RT (left), D (center) and S with order 1 (right).
It can be seen that, as expected, the RT estimation presents more local variability than D or S.
In addition, the spatial distribution of wave groups is easier to determine in D and S than in RT, due to the high variability of the envelope derived by RT, which was previously commented.
This spatial distribution of the groups is identified by the connecting areas of the sea surface where contiguous wave heights are higher than a given threshold value, which is typically consider as the significant wave height $H_s$~\cite{Ochi2005}.
As the data shown in Fig.~\ref{fig:EnvelopeFields2D} correspond to local wave height values normalized by the significant wave height $H_s$ (i.e. $\bar{H} = H / H_s$), a good threshold value to identify those connecting areas of the groups occurs when $\bar{H} \ge 1$, which correpsonds to values colored from green to red in Fig.~\ref{fig:EnvelopeFields2D}.
Furthermore, it can be seen that there are not big differences between D and S. Hence, for most practical applications D is enough to determine the spatial extension of the wave groups though the connecting areas with local heights higher than a given threshold.
Furthermore, in case of using the S method, not higher order than 1 is needed.
To illustrate the high local variability of $H_{RT}$ Fig.~\ref{fig:HeighRTSpline} shows a 3D plot of the estimation from RT and its corresponding results from S.
As in the case of the Draupner wave shown in Fig.~\ref{fig:DraupnerWave}, the S estimation has lower values and lower variability than RT.
Although the RT estimation provides high variability and high values of the local height, it is the analytical solution of the upper envelope for linear wave fields.
It is important to check if a numerical method based on the geometrical structure of the data, as D and S are, can deliver estimations of the local wave height than can follow some of the dynamical properties of the wave field.
Fig.~\ref{fig:EnvelopeSpectrakHw} shows the spectrum of the local height in the wave number ${\bf k}_H$ and frequency $\omega_H$ space derived from the spatio-temporal evolution of the local heights.
Note that the wave numbers and frequencies $({\bf k}_H, \omega_H)$ of the local height are denoted differently from the wave numbers and frequencies $({\bf k}, \omega)$ of the wave field as they are two different stochastic processes.
It can be seen that in the three cases (i.e RT, D and S) the main contribution to the spectral energy is located in the subharmonic of the dispersion relation.
The plots are shown in logarithmic scale (dB) to see the distribution of the lower values that can affect the local spatio-temporal variability of the respective estimation of the local wave height.
The smaller contributions are mainly located in a higher harmonic of the dispersion relation (dashed lines in Fig.~\ref{fig:EnvelopeSpectrakHw}).
The local envelope is related to the envelopes and the main contribution should evolve in the spatio-temporal domain with velocities close to the group velocity of linear gravity waves $c_g$ \cite{Nieto-Borge13}.
Fig.~\ref{fig:TemporalEvolution} illustrates the spatio-temporal evolution of the local envelope estimations along the mean wave propagation direction.
In those figures two lines are superimposed indicating the group velocity for the mean wave length $c_g(\lambda_m)$ and the peak wave length $c_g(\lambda_p)$, where $\lambda_m$ and $\lambda_p$ are estimated from the wave number spectrum $F({\bf k})$.
In a similar way than RT, the main contributions of the local heights derived from D and S propagate with the group velocity of the waves.
Furthermore, those estimations permit to analyze the persistence of the wave groups in time and space.

\section{Effect of measurement errors.}
\label{sc:additional_observations}

In this section we consider the effect of measurement errors. In order to do this, we first analyze this effect on the mathematical method in Section \ref{sc:MathematicalMethods}, and then we report on some simulations carried out to address this question.

\subsection{Effect on the mathematical method.}

Recall that the method in Section \ref{sc:MathematicalMethods} consists of three steps, referred to as Step 1, Step 2 and Step 3. In Step 1 measurement errors may cause a displacement of the local extrema detected by the method with respect to the real ones. This essentially affects observations whose elevations are ``very close", in the sense that the difference of the real elevations is within the range of the measurement error. However, in such a situation the configuration detected by the method is also admissible, since it does not provide a description inconsistent with the real situation. Concerning Step 2, the Delaunay triangulation only uses the grid points; even if the grid points are known with some error, the Delaunay triangulation can be computed following the exact geometric-computation paradigm using, e.g., the library CGAL, which guarantees robustness of the applied algorithms~\cite{CGALrobustness}.

In Step 3, we can analyze the effect of measurement errors by studying Eq. \eqref{splines-eq}. For simplicity, let us call ${\bf A}= {\bf \Phi}^{-1}$, ${\bf B}={\bf \Psi}^{-T}$, so that Eq. \eqref{splines-eq} is written as

\begin{equation}\label{splines-eq2}
{\bf C}={\bf A}\cdot {\bf F} \cdot {\bf B}.
\end{equation}

Measurements errors do not affect either ${\bf A}$ or ${\bf B}$, but they do affect ${\bf F}$: taking measurement errors into account, we must replace ${\bf F}$ by ${\bf F}^{\star}={\bf F}+\delta {\bf F}$, where ${\bf F}$ is the matrix containing the exact wave elevations, and $\delta {\bf F}$ contains the measurement errors in the wave elevations. Then we have
\[
{\bf A}\cdot ({\bf F}+\delta {\bf F})\cdot {\bf B}={\bf A}\cdot {\bf F}\cdot {\bf B}+{\bf A}\cdot \delta {\bf F}\cdot {\bf B}.
\]

The term ${\bf A}\cdot \delta{\bf F}\cdot {\bf B}$ evaluates the effect of measurement errors. In order to analyze it, we can compare the norms $\Vert {\bf A}\cdot \delta {\bf F}\cdot {\bf B} \Vert$ and $\Vert {\bf A}\cdot {\bf F}\cdot {\bf B}\Vert$. There are several matrix norms in the literature; the so-called  $\Vert \bullet\Vert_1$ is defined as the maximum of the sums of the absolute values of the entries in the columns of $\bullet$. Using this norm, and using also the fact that the norm of the product is bounded by the product of the norms, we get
\[
\Vert {\bf A}\cdot \delta {\bf F}\cdot {\bf B} \Vert_1 \leq \Vert {\bf A}\Vert_1 \cdot \Vert \delta {\bf F}\Vert_1\cdot \Vert {\bf B}\Vert_1.
\]
Denoting by $\epsilon$ an upper bound of the entries of $\delta {\bf F}$ (i.e. an upper bound on the measurement errors of the wave elevations), we have
\[
\Vert {\bf A}\cdot \delta {\bf F}\cdot {\bf B} \Vert_1 \leq \Vert {\bf A}\Vert_1 \cdot  \Vert {\bf B}\Vert_1 \cdot \widehat{N}\cdot \epsilon,
\]
where $\widehat{N}=\mbox{max}\{N_x,N_y\}$. On the other hand,
\[
\Vert {\bf A}\cdot {\bf F}\cdot {\bf B} \Vert_1 \leq \Vert {\bf A}\Vert_1 \cdot \Vert {\bf F}\Vert_1\cdot \Vert {\bf B}\Vert_1.
\]
Therefore, the quotient $\frac{\widehat{N}\cdot \epsilon}{\Vert {\bf F}\Vert_1}$ is a good estimator for measuring the relative error in step (3). Unless all the wave elevations are close to zero, which is not a realistic assumption, $\Vert {\bf F}\Vert_1$ is not close to zero, and this estimator, and therefore the relative error, is expected to be small.


\subsection{Simulations}

To estimate how the measurement errors of the wave elevation affect the estimation of the local wave height, different simulations have been carried out using Gaussian wave fields.
The wave elevation $\eta({\bf r})$ of each simulation was normalized as $\tilde{\eta}({\bf r}) = \eta({\bf r}) / \sigma$, where $\sigma$ is the standard deviation of the simulated wave field.
Therefore, the simulated normalized wave field $\tilde{\eta}({\bf r})$ is a zero-mean Gaussian distributed process having unit variance. With those simulated data, a white noise $n({\bf r})$ of standard deviation $\sigma_n$ has been added to the normalized wave field.
i.e. $\tilde{\eta}_n({\bf r}) = \tilde{\eta}({\bf r}) + n({\bf r})$.
In these simulations, values of $\sigma_n$ between 5\% and 15\% of the standard deviation of the simulated wave field have been used.
From those data, the estimation of the local height derived from the normalized wave field elevation by applying the proposed spline method, $\tilde{H}_D$, was compared to the corresponding local height with noise addition, $\tilde{H}_{D_n}$. Then, three expressions for the error estimation were considered:

\noindent
\begin{equation}
   \mbox{Normalized r.m.s. error:}
   \quad
   NRMS = \sqrt{\left< \left( \frac{\tilde{H}_{D_n} - \tilde{H}_D}{\tilde{H}_D} \right)^2 \right>}
   \mbox{,}
\end{equation}

\noindent
\begin{equation}
   \mbox{Normalized bias:}
   \quad
   NB = \left< \frac{\tilde{H}_{D_n} - \tilde{H}_D}{\tilde{H}_D} \right>
   \mbox{,}
\end{equation}

\noindent
\begin{equation}
   \mbox{Normalized absolute error:}
   \quad
   NAE = \left< \frac{\left| \tilde{H}_{D_n} - \tilde{H}_D \right|}{\tilde{H}_D} \right>
   \mbox{,}
\end{equation}

\noindent
where the brackets $\left< \bullet \right>$ indicate the mean value.

The obtained results are shown in Table~\ref{Table:errors}.

\begin{table}[htb]
  \centering
\begin{tabular}{| c | c | c | c |} \hline
   $\sigma_n$ & $NRMS$ & $NB$ & $NAE$ \\ \hline
   0.050 &  0.147 & -0.021 & 0.086 \\ \hline
   0.075 &  0.189 & -0.019 & 0.120 \\ \hline
   0.100 &  0.215 & -0.027 & 0.137 \\ \hline
   0.150 &  0.236 & -0.020 & 0.163 \\ \hline
\end{tabular}
\caption{Estimation of the errors in the local height.}\label{Table:errors}
\end{table}

As the local height depends on the distance between local maxima (i.e. wave crests) and closer local minima (i.e. wave troughs), a measuring error, both in crests and troughs, must contribute in the error of the local height in an order of twice the error of the wave elevations.
The estimation of the errors shown in Table~\ref{Table:errors} are consistent with this assumption.

\section{Application of the method to wave elevation map measured by an X-band radar} \label{sc:new}
\label{sc:Radar}

We have also applied our method to a wave elevation field estimated from a X-band marine radar image of the sea surface.
The radar is mounted at the German research platform of FINO~1, which is located on the North Sea (Lat. 54$^\circ$0.53'~N, Long. 6$^\circ$35.15'~E).
The measurement was taken on November, 15$^{\rm th}$, 2000 at 17:08:10 UTC.
The recorded significant wave height was $H_s = 4.1$~m.
The marine radar used for this measurement had a range resolution of 7.5~m and it was mounted about 30~m over the mean sea level.
The radar system used a commercial WaMoS-II A/D converter to sample the radar signal providing a radar image of the sea surface that is coded on 256 gray levels (i.e. one unsigned byte) \cite{Hessner+Reichert2006,Hessner2008}.
As a result of the WaMoS-II A/D conversion, the spatial resolutions ($\Delta x$, $\Delta y$) of the provided radar image have the value of the range resolution (i.e. $\Delta x = \Delta y = 7.5$~m).

\begin{figure}[ht]
\begin{center}
 \includegraphics[width=0.7\textwidth]{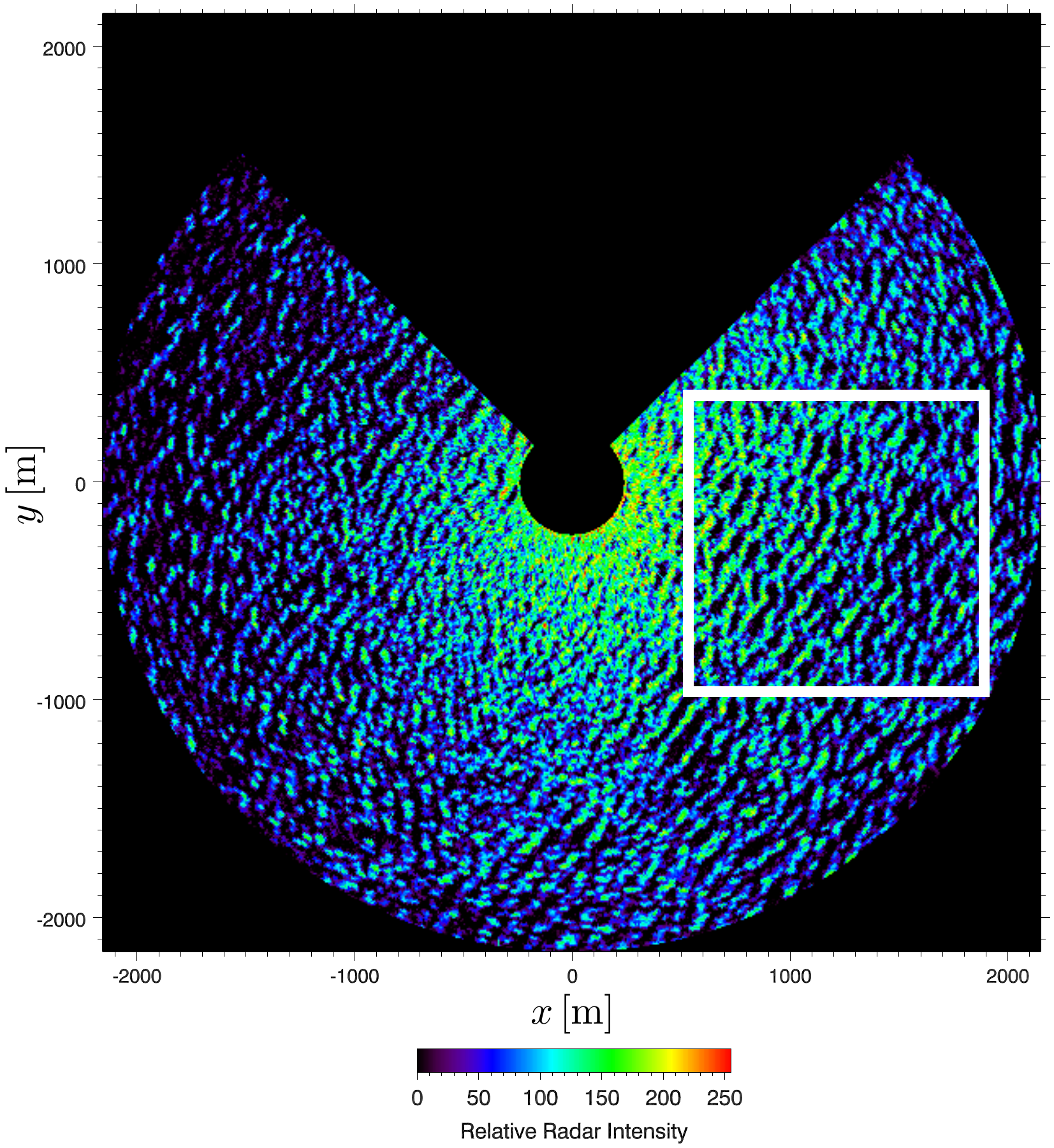}
\end{center}
\caption{X-band radar image of the sea surface taken at the FINO~1 Research platform at the North Sea. The square indicates the area where the wave inversion scheme has been applied to estimate the wave elevation field.}\label{fg:SeaClutter}
\end{figure}

\begin{figure}[ht]
\begin{center}
 \includegraphics[width=\textwidth]{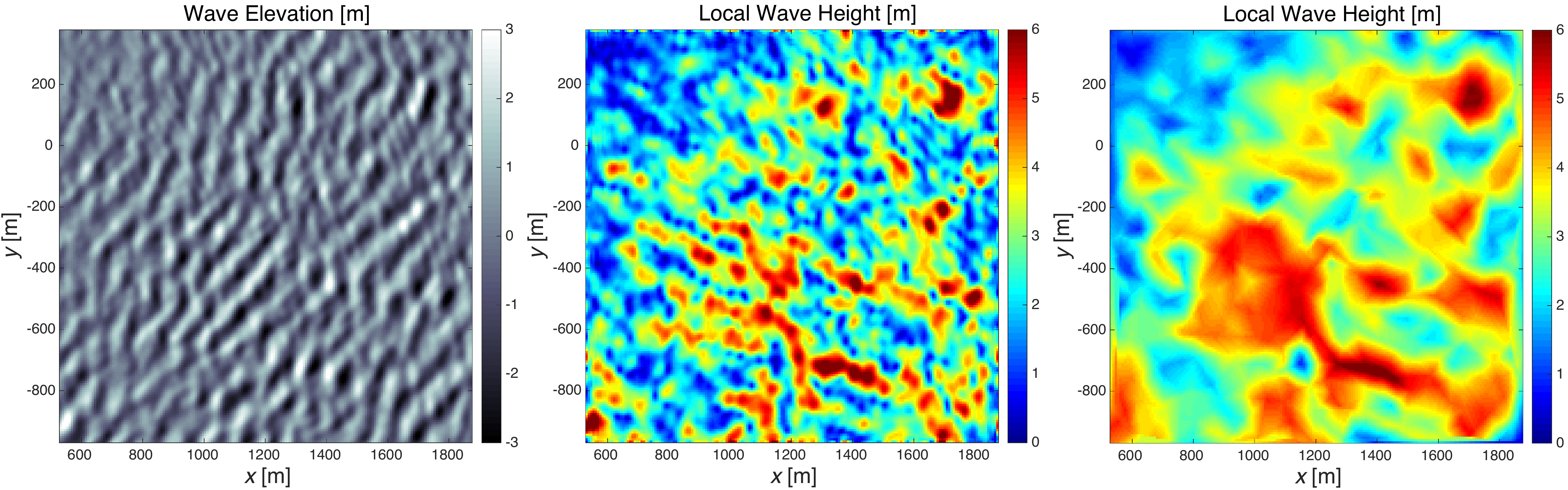}
\end{center}
\caption{Estimation of the sea surface elevation derived from the squared area shown in Figure~\ref{fg:SeaClutter} (left), corresponding local wave height derived from RT (center), and corresonding local wave height derived from D (right).}\label{fg:SeaClutterWavesHeight}
\end{figure}

Figure \ref{fg:SeaClutter} shows the corresponding radar image taken at the FINO~1 platform provided by the WaMoS-II system.
A description of this measuring system can be found in \cite{Ziemer+Dittmer1994,Ziemeretal2004,Hessner2008}.
From this radar image the wave elevation $\eta({\bf r})$ can be estimated by using inverse modeling techniques \cite{Nieto-Borge04}.
Description of the schemes to retrieve the wave elevation field can be seen in detail in \cite{Dankert+Rosenthal2004b,Nieto-Borge04,Hessner+Reichert2006}.
The left part of Figure \ref{fg:SeaClutterWavesHeight} shows the estimation of the wave elevation $\eta({\bf r})$ from the radar image illustrated on figure \ref{fg:SeaClutter}.
The estimation of the local wave height in the spatial domain has been carried out using both the RT (centered part of Figure \ref{fg:SeaClutterWavesHeight}) and D (right part of Figure \ref{fg:SeaClutterWavesHeight}) methods.
The maximum value of the local wave height derived from RT is $H_{max_{RT}} = 7.5$~m, giving a ratio with respect to the significant wave height of $H_{max_{RT}} / H_s= 1.8$~m, while the corresponding values for the D estimation are $H_{max_{D}} = 6.4$~m, $H_{max_{D}} / H_s= 1.6$~m respectively.
The higher value of the RT estimation is a consequence of the higher variability of the envelope derived from the RT method.
One can see that there is more noise in the estimation of the local wave height derived from RT than in the estimation coming from D, being the spatial structure of the groups easier to identify in the D estimation compared to the RT estimation.

\section{Conclusions and outlook}
\label{sc:Conclusions}

Estimation of individual wave heights and wave groups is commonly carried out from wave elevation time series.
The methods for analyzing individual waves in the temporal domain are difficult to generalize when the wave fields are described in the spatial (i.e., ${\bf r} = (x, y)$) or the spatio-temporal (i.e., $({\bf r}, t) = (x, y, t)$) domains, where the number of parameters to describe the variability of the wave fields is higher.
This fact is specially relevant when studying, for instance, individual waves using different imaging systems, such as microwave remote sensing techniques, which are able to scan large areas of the sea surface.

For linear wave fields, an option to analyze wave heights and groups in space and time is to use the Riesz transform (RT), which is a multidimensional generalization of the Hilbert transform, and which provides results consistent with the wave dynamics for linear gravity waves. However, the use of RT presents some problems; in particular, the high spatio-temporal variability of the envelope, and the fact that the upper and lower envelopes are always symmetric, therefore providing larger values of the estimated local wave height.
As an alternative method to estimate the non-symmetrical upper and lower envelopes, this work proposes the use of discrete and continuous techniques based on tools from Computational Geometry and Computer Aided Geometric Design. First, we use the well known Delaunay triangulation to construct a piecewise-linear model of the upper and lower wave envelopes. Then, we refine the piecewise-linear model by using tensor-product splines.

The proposed methods permit to obtain more reliable values of the local wave heights compared to RT, as shown not only in simulations, but also in some examples coming from real data. In addition, the surfaces of local wave height derived by our techniques are dynamically consistent, since the estimations of the main contribution of the wave energy, given by higher values of the local wave height, propagates with the group velocities of the wave field.

Although the proposed techniques have mostly been applied in this work to second-order non-linear stochastic wave fields, they are also suitable for any kind of spatio-temporal description of wave field, e.g. the ones derived from numerical models taking into account higher order contributions.

\section*{Acknowledgements}
The Draupner record data were kindly provided by Statoil. David Orden has been partially supported by MINECO Projects MTM2014-54207 and MTM2017-83750-P (AEI/FEDER, UE), as well as by H2020-MSCA-RISE project 734922 - CONNECT.
The WaMoS-II data at FINO~1 platform were kindly provided by Rutter Inc.
The FINO~1 platform is owned operated by \em{Bundesamt f\"ur Seeschiffahrt un Hydrographie} (BSH).
Juan G. Alc\'azar is supported by the Spanish \em{Ministerio de Econom\'{\i}a y Competitividad and by the European Regional Development Fund (ERDF), under the project  MTM2017-88796-P, and is a member of the Research Group {\sc asynacs} (Ref. {\sc ccee2011/r34}).}








\end{document}